\documentclass[letterpaper,twocolumn,10pt]{article}
\usepackage{usenix,epsfig,endnotes,multirow,longtable,multicol,tabularx,adjustbox,longtable,amssymb,subfig,subcaption}
\newcommand{{\sh}}[0]{smart home}
\newcommand{{\spa}}[0]{smart home personal assistant}
\begin{document}

\date{}

\title{Natural Language but Omitted? On the Ineffectiveness of Large Language Models' privacy policy from End-users' Perspective}

\author{
{\rm Shuning Zhang}\\
Tsinghua University
\and
{\rm Haobin Xing}\\
Tsinghua University
\and
{\rm Xin Yi}\\
Tsinghua University
\and
{\rm Hewu Li}\\
Tsinghua University
}

\maketitle

\thispagestyle{empty}

\subsection*{Abstract}

LLMs driven products were increasingly prevalent in our daily lives, With a natural language based interaction style, people may potentially leak their personal private information. Thus, privacy policy and user agreement played an important role in regulating and alerting people. However, there lacked the work examining the reading of LLM's privacy policy. Thus, we conducted the first user study to let participants read the privacy policy and user agreement with two different styles (a cursory and detailed style). We found users lack important information upon cursory reading and even detailed reading. Besides, their privacy concerns was not solved even upon detailed reading. We provided four design implications based on the findings.


\section{Introduction}

The rapid development of products based on Large Language Models (LLMs) led to their widespread adoption for various tasks. In 2023, the user base of GPT alone easily surpassed one hundred million \footnote{https://explodingtopics.com/blog/chatgpt-users}, with major corporations like Apple \footnote{https://www.theverge.com/2023/9/6/23861763/apple-ai-language-models-ajax-gpt-training-spending} and Samsung \footnote{https://techcrunch.com/2023/11/08/samsung-unveils-chatgpt-alternative-samsung-gauss-that-can-generate-text-code-and-images/} accelerating the deployment of their own LLM products and services. Key providers of LLMs include tech giants such as OpenAI \footnote{https://platform.openai.com/docs/models/overview}, Google \footnote{https://deepmind.google/technologies/gemini/}, Alibaba \footnote{https://tongyi.aliyun.com}, as well as numerous startups.

LLM-based products enable users to interact using natural language inputs and outputs, allowing for intuitive usage without the need for specialized training or learning. However, this convenience comes with a significant concern: the threat to users' privacy and security. Various articles have already pointed out new privacy and security issues unique to LLMs, stemming from their interaction and training methods.

While previous research discussed privacy and security issues related to LLM systems and prompts \cite{zhang2024s}, the privacy policies and user agreements – which are the primary means of informing users about their privacy and security – have not been thoroughly studied. These documents, which users must agree to before using the services, typically outline the handling of personal information and security protection measures, playing a crucial role in safeguarding user privacy \cite{chang2018role,wu2012effect}.

\begin{figure}[htbp]
    \subfloat[ChatGPT.]{
        \includegraphics[width=0.5\linewidth]{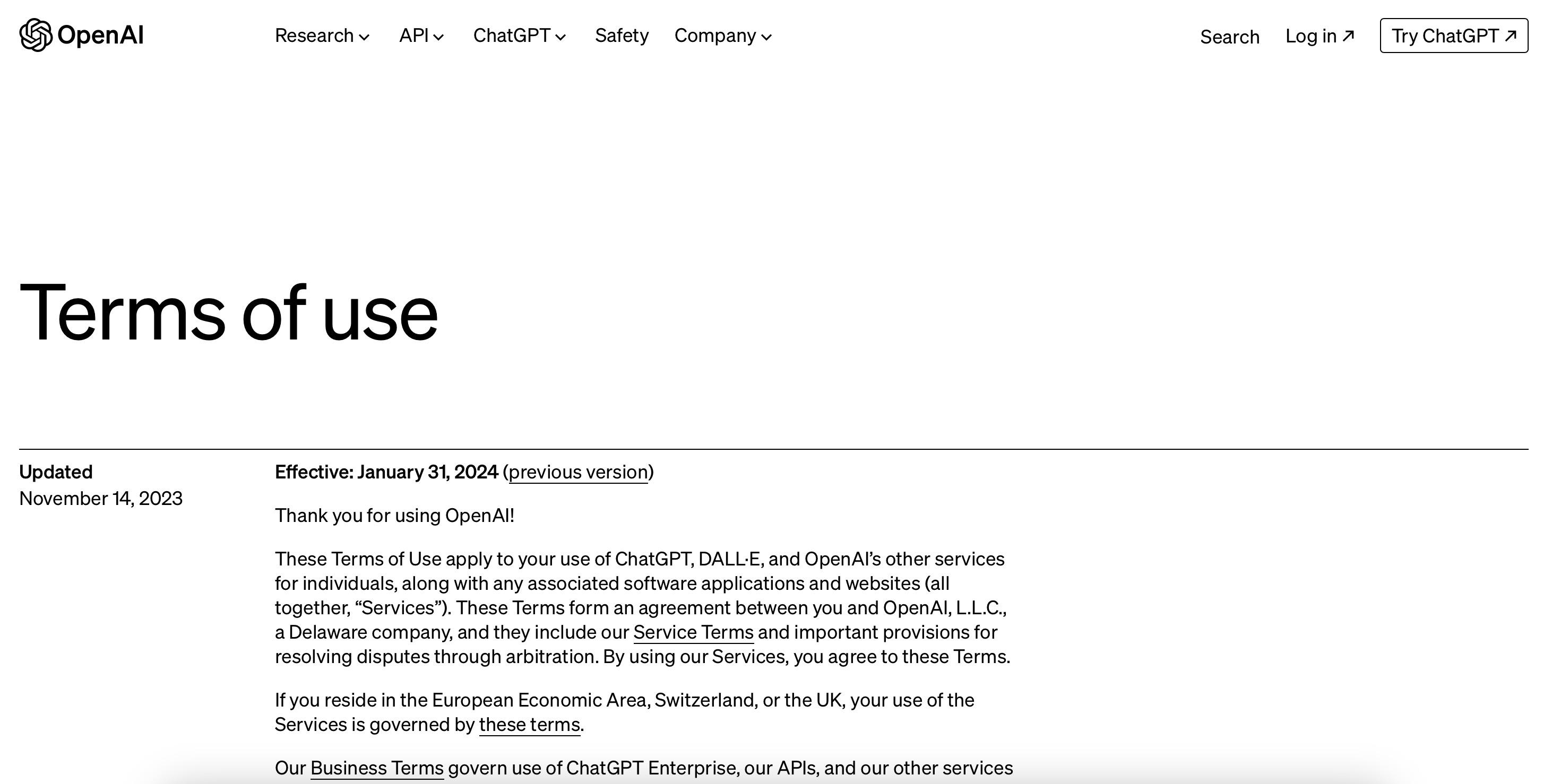}
    }
    \subfloat[Wen Xin Yi Yan.]{
        \includegraphics[width=0.5\linewidth]{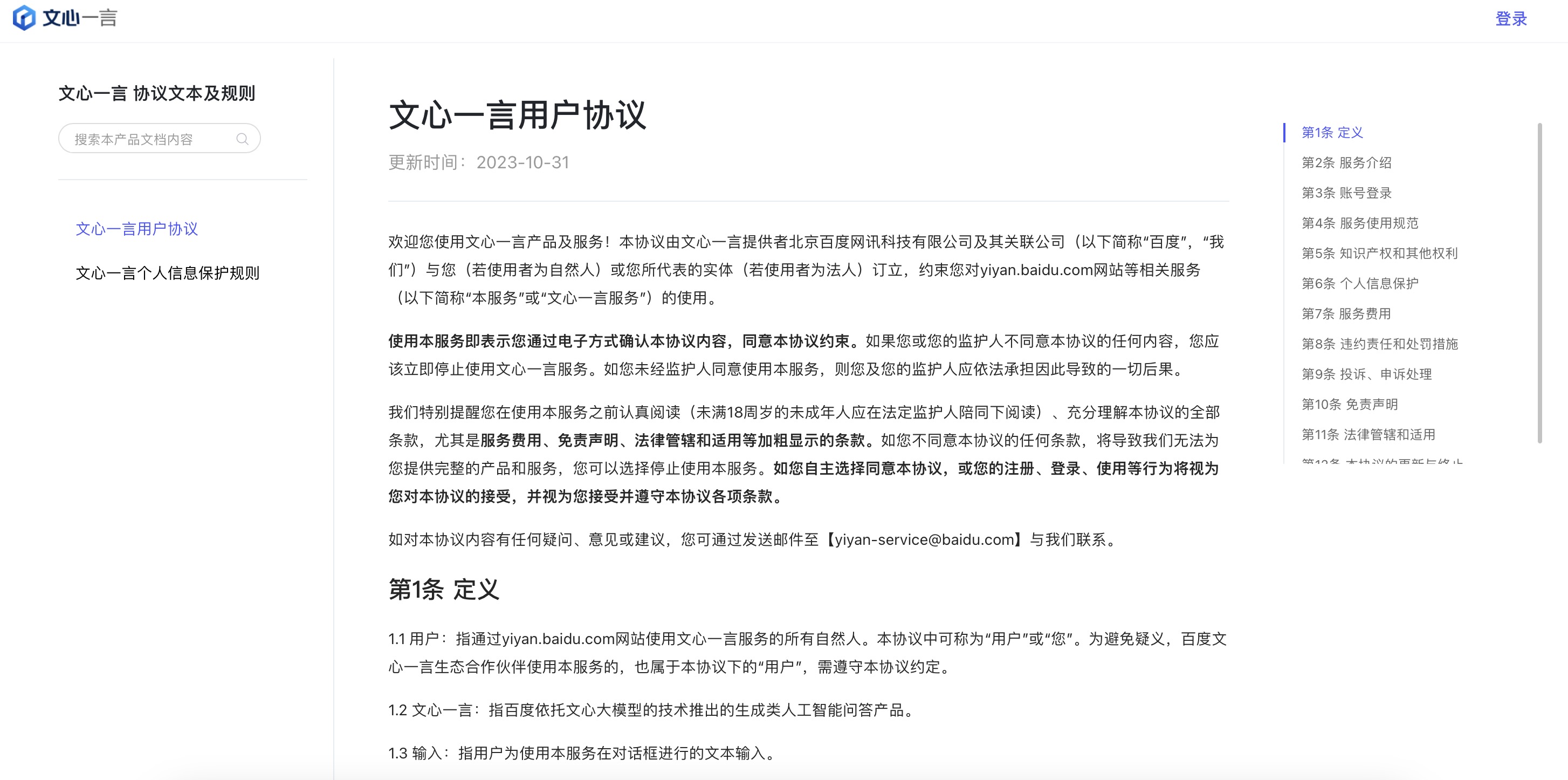}
    }

    \subfloat[PI.]{
        \includegraphics[width=0.5\linewidth]{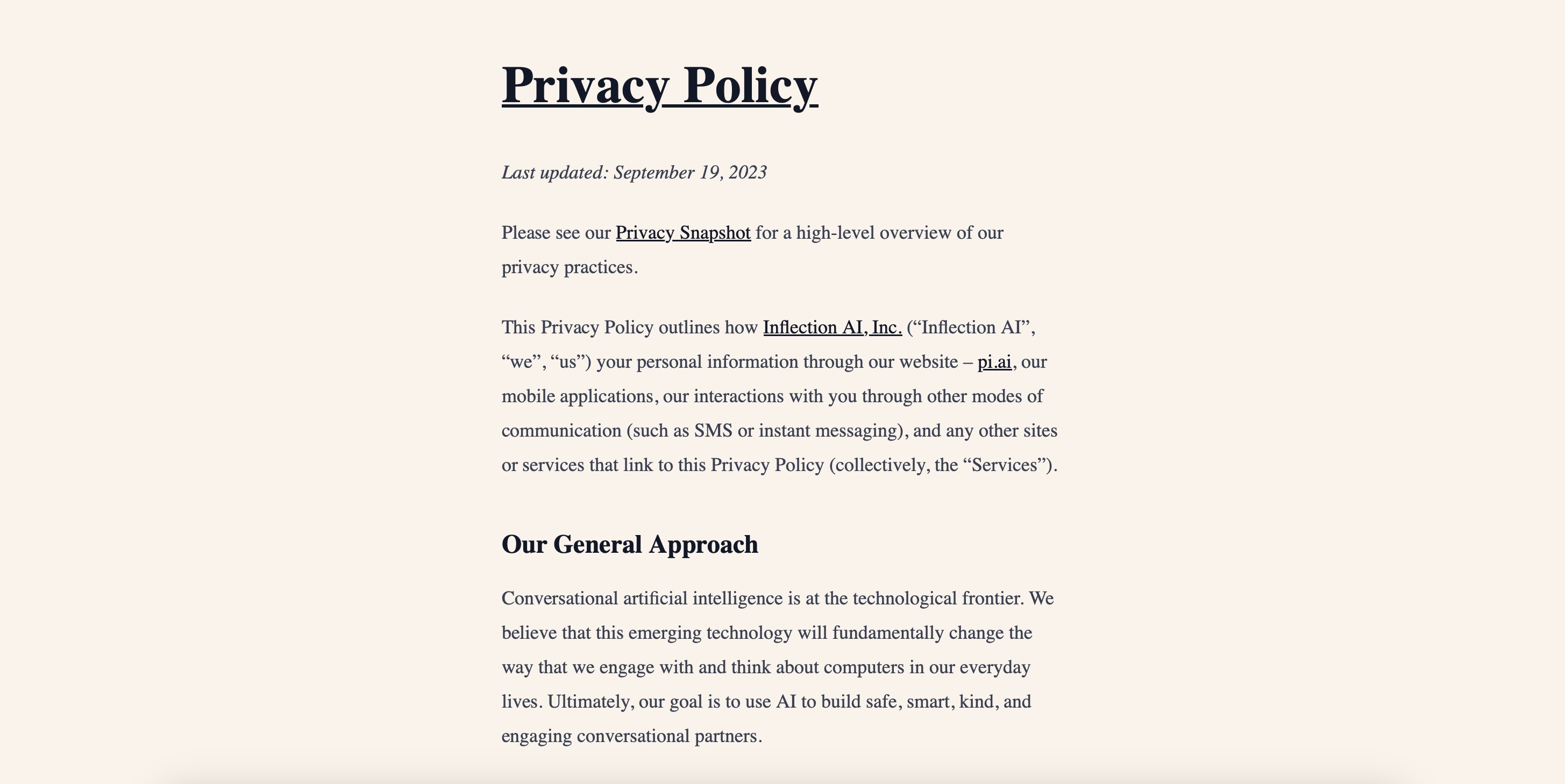}
    }
    \subfloat[Zi Dong Tai Chu.]{
        \includegraphics[width=0.5\linewidth]{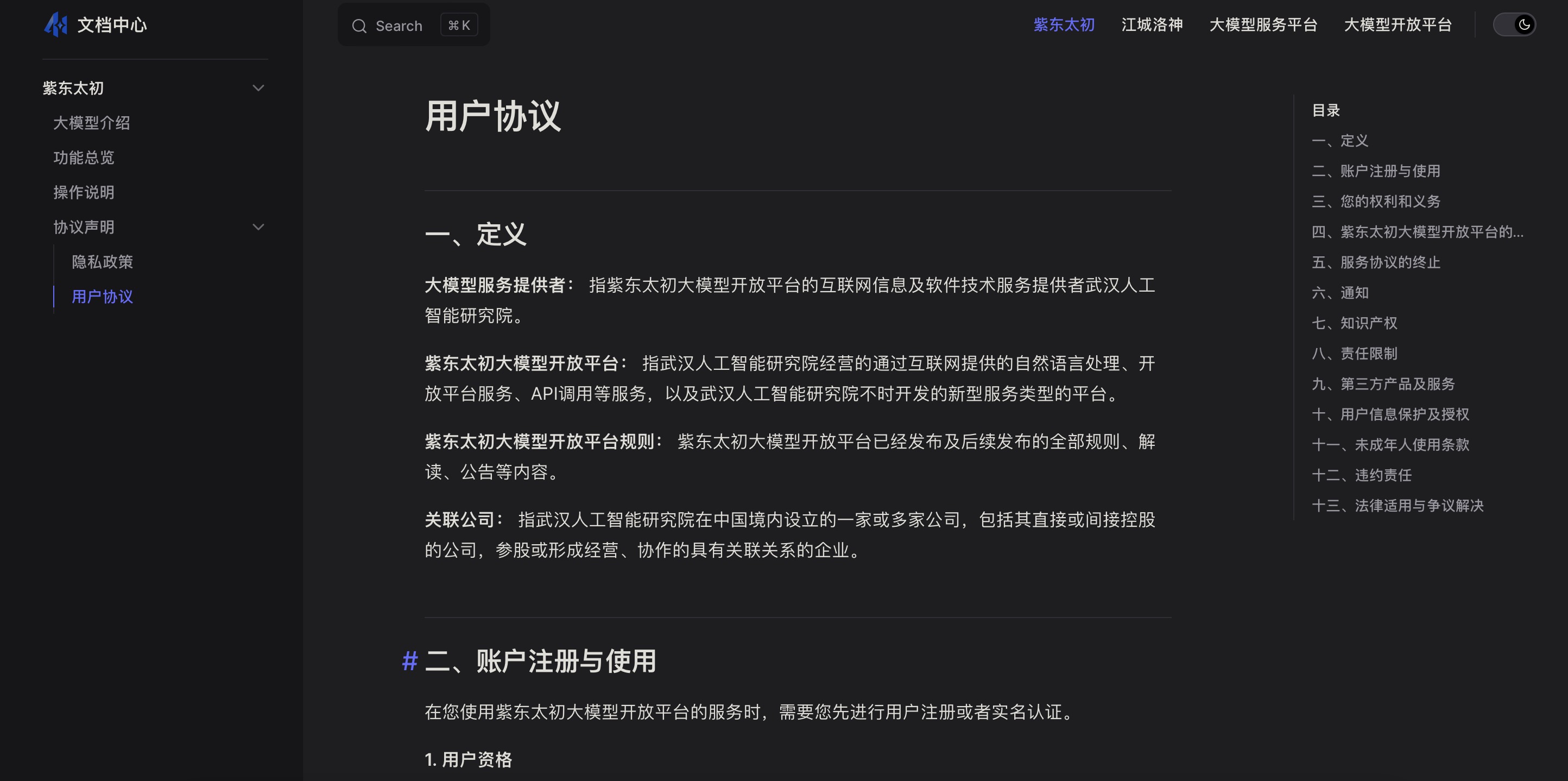}
    }
    \caption{Typical privacy policy for AI products. (a) ChatGPT. (b) Wen Xin Yi Yan. (c) PI (d) Zi Dong Tai Chu. }
\end{figure}



LLMs adopted a natural language interaction style, during which participants' personal information was more likely to be leaked and used. Besides, LLM-based products had more intellectual property, legal and usage issues. Thus, their privacy policy raised novel problems and required users' finer examination. Hence, we proposed the following research questions:

\begin{enumerate}
    \item What are the differences between the privacy policies and user agreements of LLMs and traditional AI? Which aspects are similar, and which are unique to LLMs?
    \item Do users thoroughly read the privacy policy of LLMs and specifically, which information tends to be overlooked?
    \item After a detailed reading, do users still perceive privacy threats in LLMs? Which aspects of LLM privacy policies and user agreements are considered inadequate or ineffective?
    \item Based on these discussions, how can privacy policy presentations and content be designed to bridge the gap between user understanding and the descriptions in LLMs privacy policy?
\end{enumerate}

While the previous study never touched the privacy policy and user agreement of LLMs based products, we used general use text-based products driven by LLM as an important yet rising case. We investigated users' privacy cognition and whether the notification of privacy related terms is effective. In this study, we let users read a privacy policy and user agreement extracted from commercial products. The users needed to first read the policy with a manner of daily reading. Then we asked them to read word by word. 

Through comparing these two groups and interviewing participants, we reached the following conclusions: 1) LLMs privacy policy and user agreement has over 50 important information, in which data privacy was the most important difference, 2) participants lacked important information upon cursory reading, which also was reflected in the shortening of their answer, 3) participants grasped more information upon detailed reading, however they still lacked important information. Worse still, the privacy policy and user agreement could not solve their privacy concerns. 

Finally, we presented 4 design implications gathering the viewpoint of participants and their problems: 1) there should be more clarity on data collection and usage, 2) the privacy policy should give users more control regarding whether to obey the user agreement, 3) data sharing and third-party processing should be transparent, 4) the privacy policy and user agreement should give users more education and support regarding how to use the product and protect their privacy.

In conclusion, this paper contributes to the field by:

\begin{itemize}
    \item We for the first conducted a large-scale study to explore users' perceptions and opinions on LLM products' privacy policies.
    \item We compared the privacy policy of LLM products with other AI products, highlighting the emphasis LLMs place on personal information privacy and processing.
    \item We revealed the ineffectiveness of users' reading and the presentation of LLMs' privacy policies and agreements, indicating a need for improvement in these areas.
    \item We proposed four design implications and improvement suggestions, further validated through interviews, focusing on visualization, content presentation, focal points, and user experience optimization.
\end{itemize}





\section{Background \& Related Work}

\subsection{LLM's Privacy Concerns}
LLM was designed for general information output given natural language input. Its surging ability demonstrated from GPT3 \cite{brown2020language} enabled that it could complete various downstream tasks. With the increasing ability of LLM (especially for GPT3.5, GPT4, LLaMa, etc.), it is increasingly adopted to achieve task automation \cite{wen2023empowering}, question answering \cite{zhuang2023toolqa}, content generation \cite{agossah2023llm}, task planning \cite{ruan2023tptu}, reasoning \cite{wang2023can} and other various tasks. Numerous researchers investigated the multi-modal capability of LLMs, including scene / picture generation \cite{zhao2023making} (e.g., DALL-E, CogVLM), video generation \cite{lin2023videodirectorgpt,huang2023free} (e.g., VideoDirectorGPT), voice interaction \cite{dong2023towards,le2023voicebox} (e.g., VoiceBox). However, most of the LLMs still adopted the default text modality for generation. Besides, most AI products powered by LLMs (e.g., CoPilot \footnote{https://copilot.microsoft.com}, ChatPaper \footnote{https://chatwithpaper.org}) adopted the text modality for interaction. Thus we chose text-based LLMs as the research target.

Accompanying the advance of LLMs' technologies, they faced privacy and security issues, in which privacy issues were mainly connected with personal data leakage. In fact, Italy has became the first Western country to temporarily block ChatGPT in April 2023 due to privacy concerns and the lack of proper regulation.\footnote{https://www.bbc.com/news/technology-65139406}. They also were hotly discussing the risks and mitigation. Researchers pointed out LLMs may memorize training data \cite{smith2023identifying,khowaja2023chatgpt}, resulting in membership inference \cite{duan2023privacy}, data extraction \cite{pan2020privacy} and model inversion \cite{pan2020privacy}. Besides, the training data and generated results may have copyrights problems \cite{khowaja2023chatgpt}. Regarding sensitive application scenarios including medical health, the sensitive data as well as the scenarios themselves has privacy risks \cite{mesko2023imperative}.



Besides privacy problems, ChatGPT also had many security problems. Gupta et al. \cite{gupta2023chatgpt} proved that Generative AI faced numerous attacks like Jailbreeaks, reverse psychology and prompt injection attacks, which resulted in security problems. Besides, LLMs potentially faced creepy use \cite{bommasani2021opportunities}, misinformation \cite{bommasani2021opportunities,weidinger2021ethical} and data poisoning \cite{bommasani2021opportunities}, discrimination \cite{weidinger2021ethical}, etc., which also affected the downstream usage.

Others proposed measurements, countermeasures and privacy protection mechanisms. Measurements showed with larger model size, more privacy information may be leaked \cite{plant2022you}. Regarding the protection, Kandpal et al. \cite{kandpal2022deduplicating} showed data de-duplication methods ensured LLMs to become more secure against privacy attacks.  Mireshghallah et al. \cite{mireshghallah2021privacy} introduced two privacy-preserving regularization methods with discriminators and a novel triplet loss term for training language models, achieving joint optimization of utility and privacy.

Still others argued the ineffectiveness of potential protection mechanisms. Brown et al. \cite{brown2022does} discusses the mismatch between popular data protection techniques and the broadness of natural language and privacy as a social norm. The existing protection methods, in their view, cannot guarantee a generic and meaningful notion of privacy for language models. The author concludes that language models should be trained on text data explicitly produced for public use. Wu et al. \cite{wu2023unveiling} also commented that although there are privacy protection mechanisms in ChatGPT blocked the access to personal data about individuals, privacy data may still leak. This inspired our work to investigate LLMs' privacy policy and user agreements. 

\subsection{Privacy Policy of AIs}
The discussion and analysis of privacy policy started from traditional AI products. Elliott et al. \cite{elliott2019culture} discussed the controversy of privacy policy and usage in AI technologies' adoption. Contissa et al. \cite{contissa2018claudette} proposed a supervised machine learning methodology to automate the evaluation of privacy policies. Focusing on voice assistants, Liao et al. \cite{liao2020measuring} measured the effectiveness of privacy policies provided by voice app developers on voice assistant platforms. By analyzing 64,720 Amazon Alexa skills and 16,002 Google Assistant actions, it was found that among the 17,952 skills and 9,955 actions with privacy policies, many of them had incorrect privacy policy URLs or broken links. Xiao et al. \cite{xiao2022lalaine} introduced Lalaine and analyzed the privacy labels of 5102 iOS applications. They revealed the prevalence and severity of privacy label non-compliance, providing insights for improving privacy label design and compliance requirements. Koch et al. \cite{koch2023ok} analyzed 3,006 privacy consent dialogs in mobile applications and found that 22.3\% of the apps have some form of dialogs, but only 11.9\% provide users with actionable choices. The majority of these apps coerce users into consent and do not offer choices. Du et al. \cite{du2023withdrawing} studied users' privacy withdrawal decisions and implemented MOWCHECKER, identifying 157 inconsistencies in third-party data collection by mobile applications. They received 23 responses, with 2 of them already fixed. Mohamed et al. \cite{mohamedattention} studied Apple's App Tracking Transparency (ATT) guidelines and identified four patterns violating ATT standards and further developed ATTCLS. By applying ATTCLS to 4,000 iOS apps, they revealed that 59\% of the alerts used four improper design patterns: misleading users, incentivizing tracking, including confusing terminology, or omitting the purpose of ATT permissions. 

With the advance of LLMs, researchers focused on using LLMs to analyze privacy policy. Tang et al. \cite{tang2023policygpt} developed PolicyGPT, a Large Language Model-based framework for analyzing privacy policies, which demonstrated high accuracy in classifying policy texts from websites and mobile applications. However, all previous work did not consider the complexity of LLMs' privacy policy and user agreement, which contained information about the personal information privacy risks and intellectual property. How users understood the privacy policys affected providers and users' profit. Thus we initiated this first study.

\subsection{Reading of Privacy Policy}
According to the US Legal Dictionary, a privacy policy in the online context is related to as ``a statement that declares a firm's or website's policy on collecting and releasing information about a visitor'' \cite{jensen2005privacy,pollach2007s,fabian2017large}. The past work regarding privacy policy consisted of: analyzing users' perception of the privacy policy \cite{ibdah2021should,mcdonald2008cost,graber2002reading}, facilitating reading and comprehension of privacy policy \cite{cranor2002web,reeder2008user,pan2023toward,kelley2009nutrition,yu2015autoppg}, automatic extracting and analyzing privacy policy \cite{stamey2009automatically,cranor2002web,tesfay2018privacyguide,rosen2013appprofiler}.

Regarding users' perception, researchers have already demonstrated that privacy policy was hard to read \cite{ibdah2021should,mcdonald2008cost,graber2002reading} and comprehend \cite{vu2007users}, due to its lengthy nature \cite{mcdonald2008cost} and complicated words \cite{fabian2017large}. Past researchers proposed various objective (e.g., RIX \cite{anderson1983lix}, LIX \cite{anderson1983lix}, FKG \cite{shedlosky2009tools}, ARI \cite{shedlosky2009tools}, FRES \cite{jensen2004privacy,flesch1949art,shedlosky2009tools,fabian2017large}, to name a few) and subjective metrics (e.g., ) for evaluting privacy policy. The objective metrics were proved highly correlated \cite{fabian2017large} and biased as they mainly used word length and difficulty to compose the metric \cite{fabian2017large,ermakova2014privacy,stevens1992measuring}. The subjective metrics were divided into two categories: cloze \cite{stevens1992measuring,singh2011user,dubay2004principles} and entries \cite{ermakova2014privacy,milne2004strategies,bansal2008moderating,bansal2008efficacy}. Cloze test were advocated \cite{stevens1992measuring,dubay2004principles}, developed \cite{stevens1992measuring,singh2011user} and proved to be effective in testing users' comprehension \cite{singh2011user}. However, it could only measure users' comprehension of the original privacy policy, without generalization to other variable forms of visualizations such as privacy label \cite{kelley2009nutrition}, thus limited in its usage scope \cite{ravichander2019challenges}. Researchers also developed subjective entries to measure users' comprehension \cite{milne2004strategies}, satisfaction \cite{milne2004strategies,mckinney2002measurement} and trust \cite{mcknight2002developing,milne2004strategies,model2003trust} towards the privacy policy statement. This work adopted these entries to evaluate users' opinions from an unbiased manner.

Owing to the hindrance of reading privacy policy \cite{graber2002reading,vu2007users}, researchers hope to facilitate privacy policy's demonstration and reading \cite{cranor2002web,reeder2008user,pan2023toward,kelley2009nutrition,yu2015autoppg}. Early efforts include Platform of Privacy Policy (P3P) framework, which extracted the privacy policy as XML files composed of data types, recipients and transmission protocols \cite{cranor2002web}. There were efforts visualizing the P3P \cite{reeder2008user}, however failed because users think P3P's visualization has 1) no focal point, 2) multiple statements, 3) confusing icons. It is deprecated since 2006 \cite{reeder2008user}. Researchers also mimicked the nutrition label to proposed privacy nutrition label \cite{kelley2009nutrition} and optimized its content \cite{pan2023toward}. Privacy label demonstrates private information collection \cite{kelley2009nutrition,pan2023toward} and fits Android \cite{yu2015autoppg}, iOS platforms \cite{zhang2022usable}. However, it is limited in the information capacity \cite{ammar2012automatic} and could not represent other important information in privacy policy such as recipients \cite{cranor2002web} and transmission protocol \cite{cranor2002web}. 

Apart from the above efforts, there were also works to automate the analyzing of privacy policies, to 1) disclose its illy-presented manner \cite{stamey2009automatically}, 2) advocate for better protection practice \cite{cranor2002web,tesfay2018privacyguide,rosen2013appprofiler}. Early adoptions included P3P \cite{cranor2002web} and Terms of Service, Didn't Read (ToS;DR) \cite{tesfay2018privacyguide}, however were limited in their generalize-ability. Later research transferred to automate the privacy policy extraction leveraging NLP techniques. Privacy Informer \cite{miao2014privacyinformer} proposed an automatic privacy policy extraction method and AppProfiler \cite{rosen2013appprofiler} manually mapped knowledge base to behavior. Usable Privacy Policy \cite{sadeh2013usable} made the first attempt on NLP based privacy policy analysis, exploring the possibility of privacy policy classification. Costante et al. \cite{costante2012websites} investigated the adaptability of rule-based and machine-learning based techniques in privacy policy analysis. PrivacyCheck \cite{zaeem2018privacycheck} leveraged data mining, achieving automatic privacy policy extraction and risk analysis. It is demonstrated through Chrome plugins, which answered basic questions about users' data privacy. Stamey and Rossi \cite{stamey2009automatically} developed Hermes, which could identify latent semantic correlation and analyzed the ambiguity of privacy policy. Privee \cite{zimmeck2014privee} is another automatic framework ensembling rules and machine learning techniques, aiming at increasing the transparency of privacy policy. These automatic extraction framework is inspiring for our work but limited in the generaliz-ability. Our methods, on the opposite leveraged the zero-shot capability of Large Language Models (LLMs) to enable privacy policy analysis without the need of training data.

\section{Motivation and Problem Definition}
\begin{figure}
    \includegraphics[width=\linewidth]{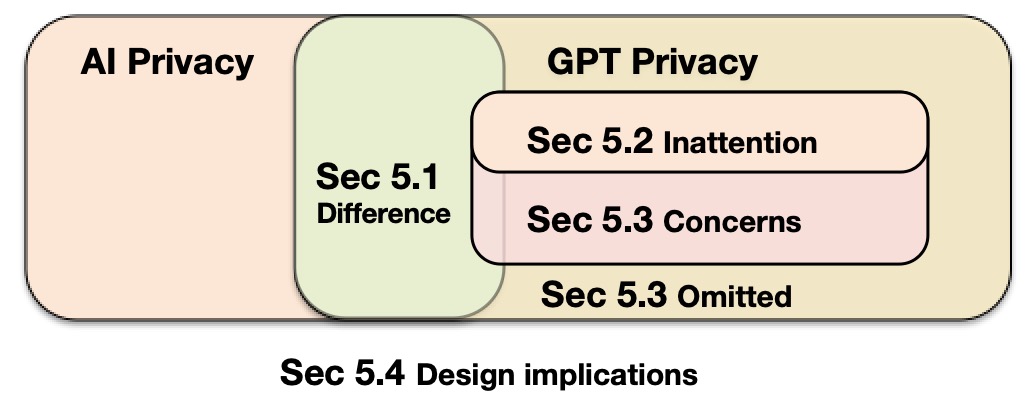}
    \caption{The problem definition and research framework of this paper.}
\end{figure}

\subsection{Motivation}
The privacy characteristics of LLMs differ from those of traditional AI products in several ways\cite{zhang2024s}. Firstly, LLMs are characterized by their extensive scale, requiring vast training corpora and often operating in a continuous training mode. Companies deploying these models have a strong incentive to utilize user-generated data for ongoing model refinement, potentially involving the use of users' privacy information.

Furthermore, users directly interact with LLMs using natural language, even though commercial LLMs often incorporate some pre-processing for front-end interactions. Essentially, users' input is generally transmitted to the LLMs without substantial alteration. In contrast, when humans provide information in natural language, they typically convey more detailed linguistic information than just clicking GUI buttons, potentially exposing more privacy-sensitive data. Therefore, users may unknowingly expose more privacy through these interactions.

Additionally, unlike traditional AI products, users of LLMs are typically less conscious of privacy concerns during their usage. Users might not be aware that their privacy information has been leaked or is at risk of being disclosed, especially when they focused on the specific questions they are asking, some of which may pertain to sensitive information, such as medical queries. Traditional AI products, in contrast, often create a strong sense of privacy awareness when accessing or providing sensitive information, exacerbating the potential privacy threats posed by LLMs.

From another perspective, the applications of large models are becoming increasingly widespread. Within a very short time, OPENAI's user base has grown to over a billion users. Even before the launch of OPENAI's ChatGPT, there were various dialogue systems and AI products based on a variety of language models. These products are rapidly transitioning to using large-scale pre-trained models as their foundation, potentially introducing additional security threats. Moreover, following the launch of OPENAI's ChatGPT, there have been downstream products using large models in China, Europe, and the United States, adding to the competitive landscape. Many of these products are beginning to offer general or specialized functionalities based on large-scale pre-trained models. As a result, users are indeed facing more severe privacy challenges in practice. 

\subsection{Problem Definition}
Focusing on the user agreement and privacy policys of LLMs, we were interested in the following research questions:

RQ1: What is the content of LLMs' privacy policy and user agreement? What is the difference between AI and LLMs' privacy policy and user agreement?

RQ2: What is participants' daily practice of viewing LLMs' and AIs' privacy policy and user agreement? Would participants have concerns?

RQ3: What would participants concentrate on when asked to view LLMs' and AIs' privacy policy and user agreement meticulously? Would participants still have concerns about their privacy risk?

RQ4: What could be done by manufacturers to improve the design of privacy policy and user agreement?

To answer these research questions, we first classified the content of user agreement and privacy policy, deriving the structure. Following this, we conducted a study to investigate: 1) the difference between LLMs' and AIs' privacy policys, 2) the inattention of participants upon a cursory reading of LLMs' privacy policys, 3) the privacy concerns and omitted content during the detailed reading towards LLMs' privacy policys. 

After finding the difference, we hoped to unveil the design implications regarding what to do to reduce participants' privacy concerns and increase their awareness as well as satisfaction. 

\section{Methodology}

\subsection{Study Design}




Our primary objective was to investigate users' comprehension of user agreements and privacy policies during the reading process. Therefore, the experimental variable revolved around \textbf{the degree of thoroughness} with which users read these documents. We established two different levels: users' typical reading speed (during which we observed that all users generally read very cursorily) and a more thorough reading speed. The criteria we examined included a $2 \times 2$ scale: the content they remember (without referring to the original text), the content they considered important (without referring to the original text), the content they remember (referring to the original text), the content they considered important (referring to the original text). We further let participants to assess the effectiveness of the user agreements and privacy policy.

Since there are numerous products based on LLMs with text input and output (e.g., ChatGPT\footnote{https://openai.com/chatgpt/, accessed by 16th Jun, 2024}, Claude\footnote{}, PI\footnote{}), we selected the most common and widely recognized models in China and the United States. As the experiment was conducted in China, our selection was based on the prevalence, frequency of use, and familiarity among Chinese users.

In order to ensure that our results have broader relevance and guidance for ordinary users, as well as for producers and developers, rather than catering to specific demographics, the criteria on the products are as follows: they needed to exhibit universality, being capable of addressing questions across various domains and designed for general conversation rather than exclusively resolving specialized questions in specific fields such as mathematics, geography, or astronomy. As a first attempt, such a selection better reflects and addresses the common issues in privacy policies of LLMs.

Given that for different LLM-based products, most privacy policy and user agreement were two separate document. Besides, we recognized that users may have distinct interpretations and understandings of these two sections. Therefore, the experiment was designed to let users to read and engage with both the user agreement and the privacy policy separately.

\subsection{Participants}
For the survey, we recruited 27 participants from the campus through snowball sampling (with an age of 22.70, SD=2.67). In order to ensure the quality of participants, we established criteria to ensure that all participants were LLMs' users. We found that participants had an average level of knowledge about LLM at 4.10, with 23 individuals using LLM products more than 100 times per year. However, the average level of understanding of LLM's privacy policy among participants was only 2.87 and the average time spent reading the privacy policy was only 4.89 minutes. The demographics of the survey were shown in Table~\ref{formal_background} in the appendix.

\subsection{Procedure}

First, we familiarized users with the experimental process and utilized a questionnaire to inquire about participants' usage experiences with LLMs and their overall perspectives on the privacy and security of LLM-based products.

Next, we employed a random selection process (using Python's random functions) to present users with the user agreement of a downstream AI product powered by a large-scale pre-trained model. Users were asked to perform an initial, cursory reading of the respective user agreement, followed by a series of questions. Subsequently, users were instructed to engage in a detailed reading of the same user agreement, accompanied by a series of follow-up questions. This marked the completion of the first experimental phase.

In the second phase, users were initially tasked with a cursory reading of the corresponding privacy policy, followed by a set of questions. Subsequently, users were instructed to engage in a detailed reading of the privacy policy, accompanied by a series of follow-up questions. This concluded the second experimental phase.

Finally, we inquired about users' privacy considerations during the design process and asked how they would design these documents if given the opportunity. The overall duration of the experiment was approximately one hour.

\subsection{Analysis Methods}



We employed a combined approach of qualitative and quantitative analysis. For qualitative analysis, we adopted thematic analysis. In our study, all interviews were recorded and transcribed verbatim. One researcher conducted transcript cleaning, removing irrelevant content such as repeated words and ineffective pauses to ensure data quality. Following transcription, we proceeded with coding the participants' responses. The coding process involved two stages: open coding and axial coding. In the initial stage, the lead researcher meticulously read through each interview transcript, identifying key phrases and statements, which led to the generation of preliminary codes. These codes represented fundamental concepts and ideas related to users' privacy awareness.

In the second stage axial coding, the lead researcher organized these preliminary codes into broader categories and sought connections among them. This process primarily entailed merging, splitting, and reorganizing codes to form more comprehensive themes. We used the categories derived from axial coding to further engage in thematic generation, refining and defining the themes. This step involved consolidating similar categories to ensure each theme was unique and representative. The aim of thematic generation was to capture the core concepts and patterns of user cognition in the interview data.

Additionally, to validate the reliability of the themes, we invited other researchers from the laboratory to independently review the coding and thematic generation processes. We also utilized selected portions of the interview data to illustrate the themes, ensuring consistency between the themes and the original data. We would report all the themes generated throughout the process along with their frequencies.

For the quantitative analysis, we initially conducted a normality test on the data. For data that met the assumptions of normality, we proceeded to perform sphericity and homogeneity of variance tests and employed the corresponding analysis of variance methods (repeated measures ANOVA) for further analysis. For most data that did not adhere to a normal distribution, we conducted non-parametric Wilcoxon test (or Friedman test) for between-group comparisons and used Nemenyi post hoc analysis for paired comparison if needed.

\section{Results}

\begin{figure*}[!htbp]
    \subfloat[]{
        \includegraphics[width=0.30\linewidth]{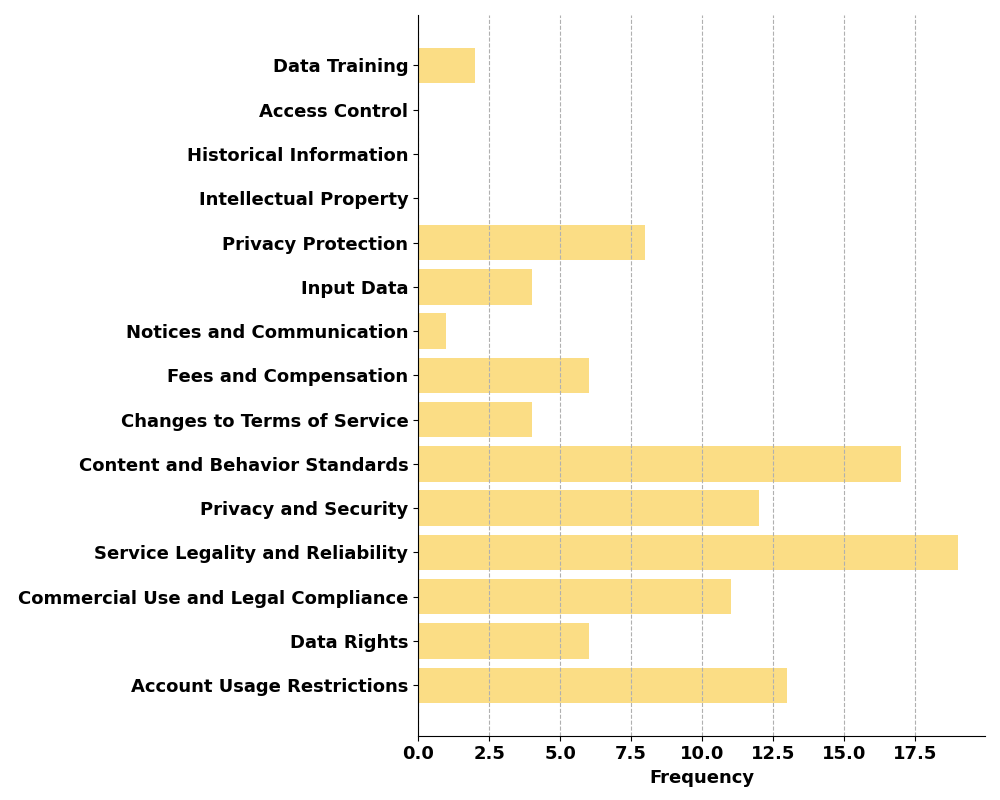}
    }
    \subfloat[]{
        \includegraphics[width=0.30\linewidth]{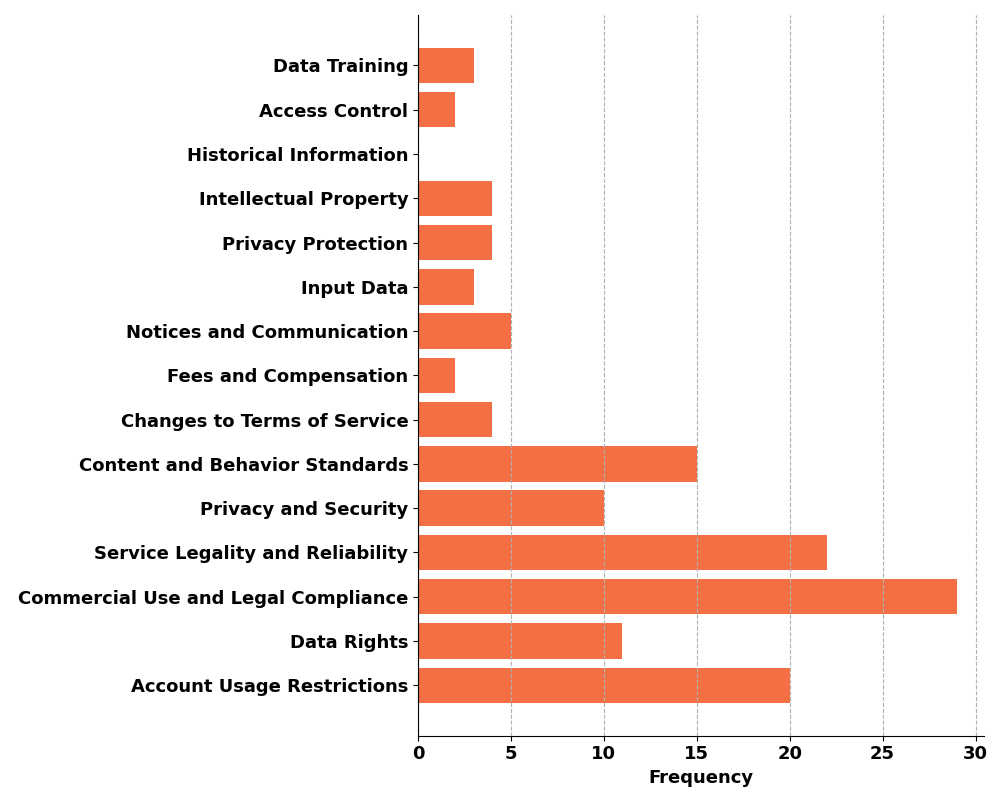}
    }
    \subfloat[]{
        \includegraphics[width=0.30\linewidth]{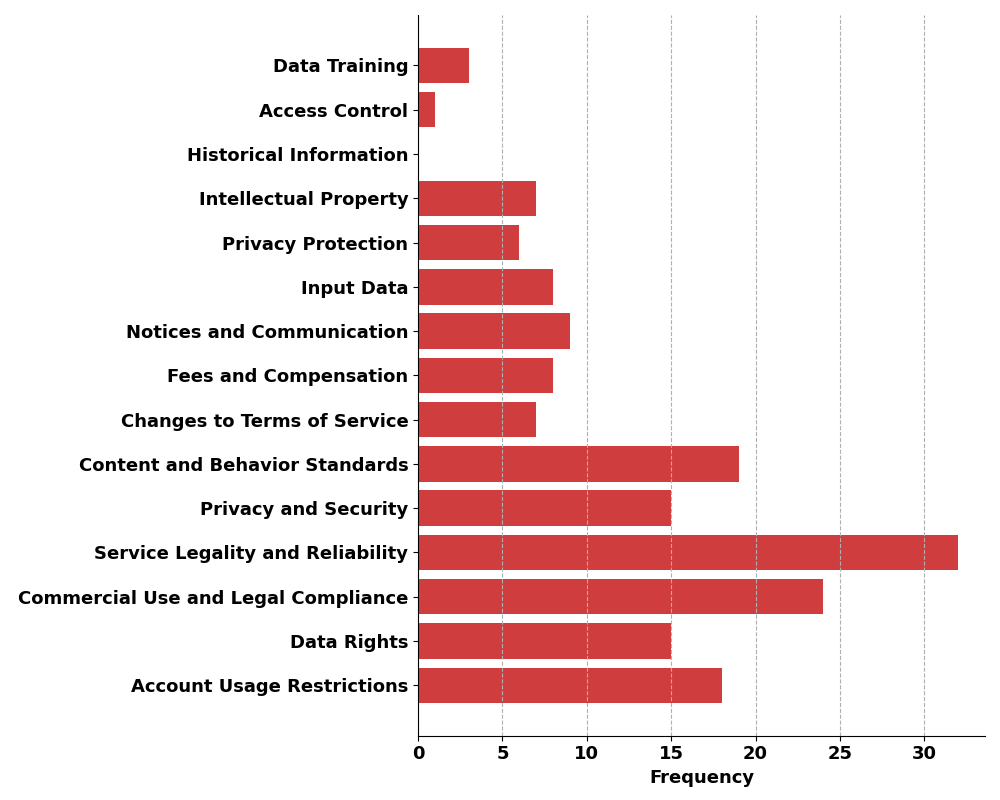}
    }
    
    \subfloat[]{
        \includegraphics[width=0.30\linewidth]{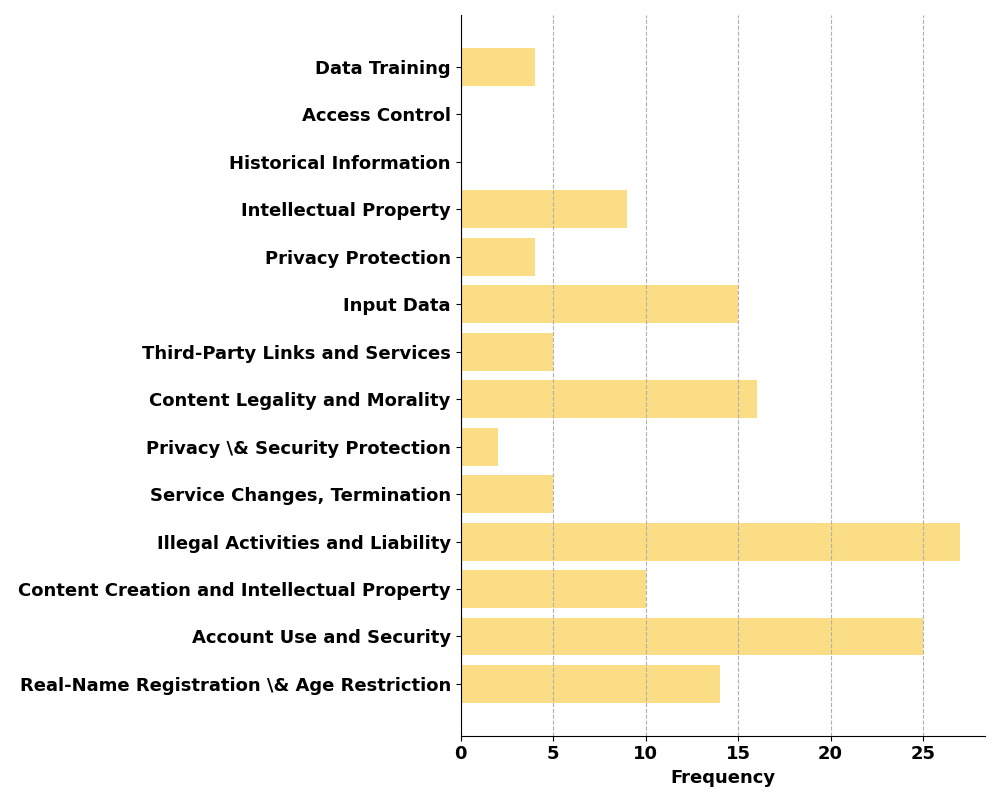}
    }
    \subfloat[]{
        \includegraphics[width=0.30\linewidth]{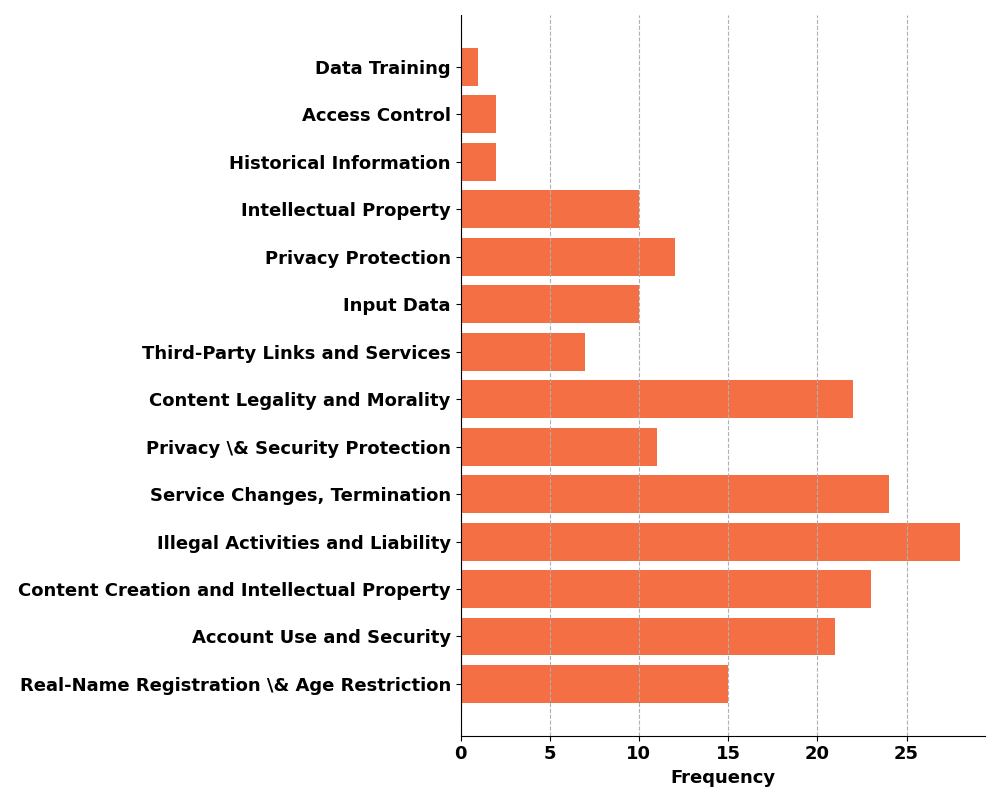}
    }
    \subfloat[]{
        \includegraphics[width=0.30\linewidth]{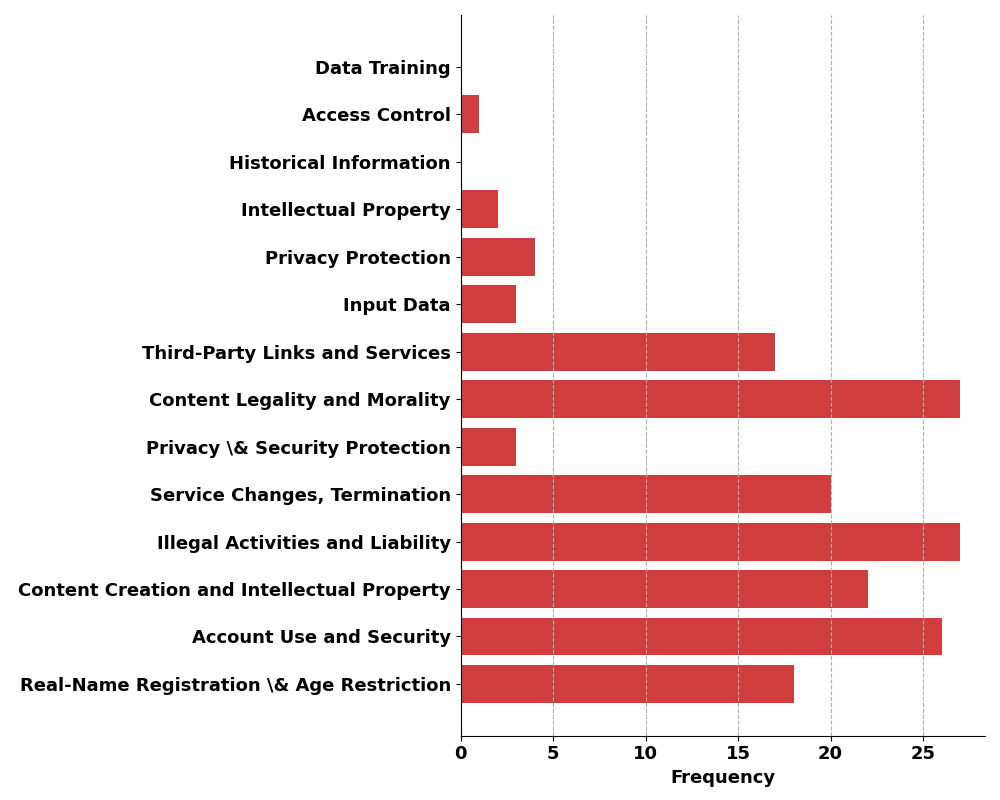}
    }

    \subfloat[]{
        \includegraphics[width=0.30\linewidth]{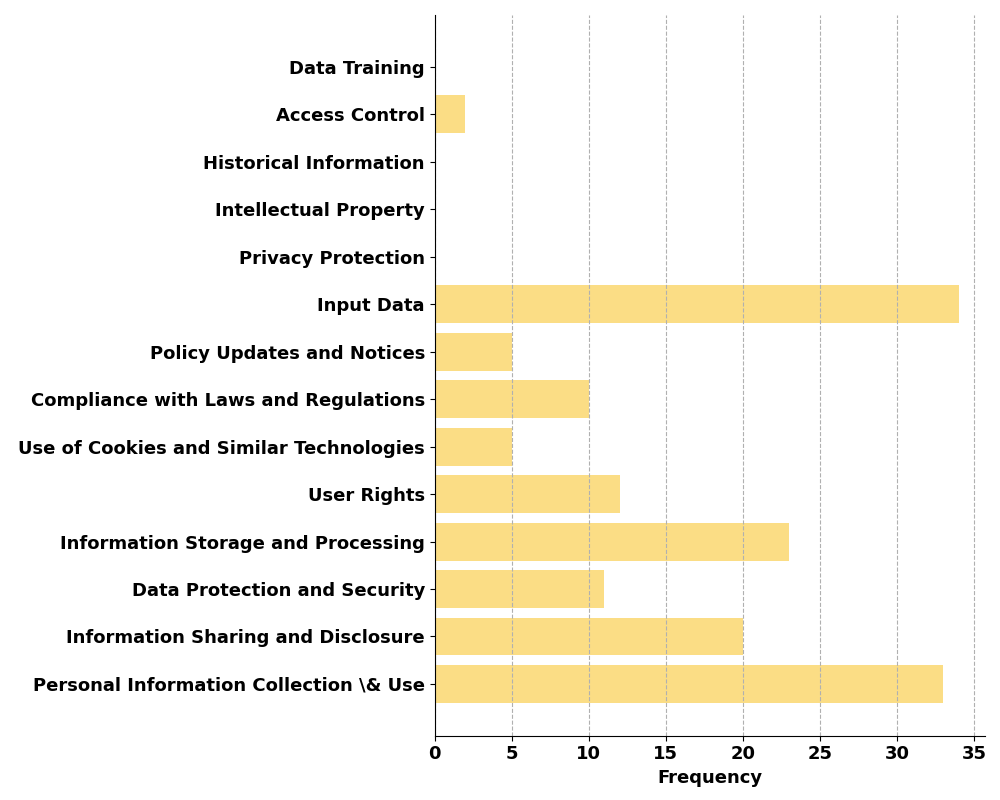}
    }
    \subfloat[]{
        \includegraphics[width=0.30\linewidth]{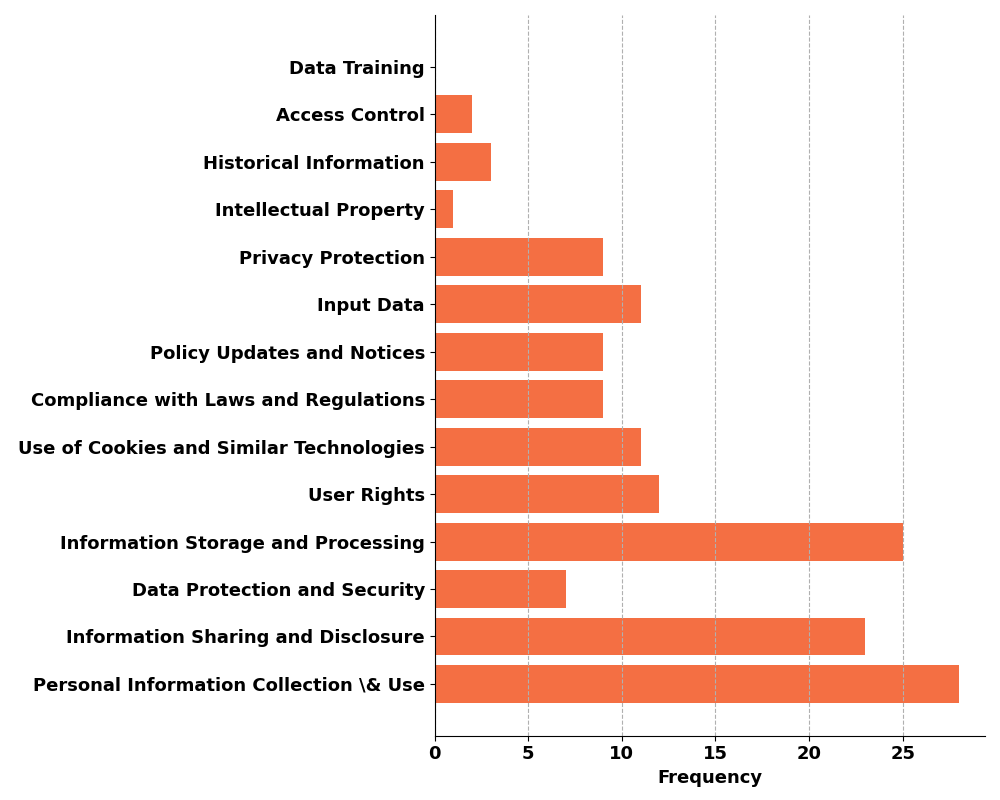}
    }
    \subfloat[]{
        \includegraphics[width=0.30\linewidth]{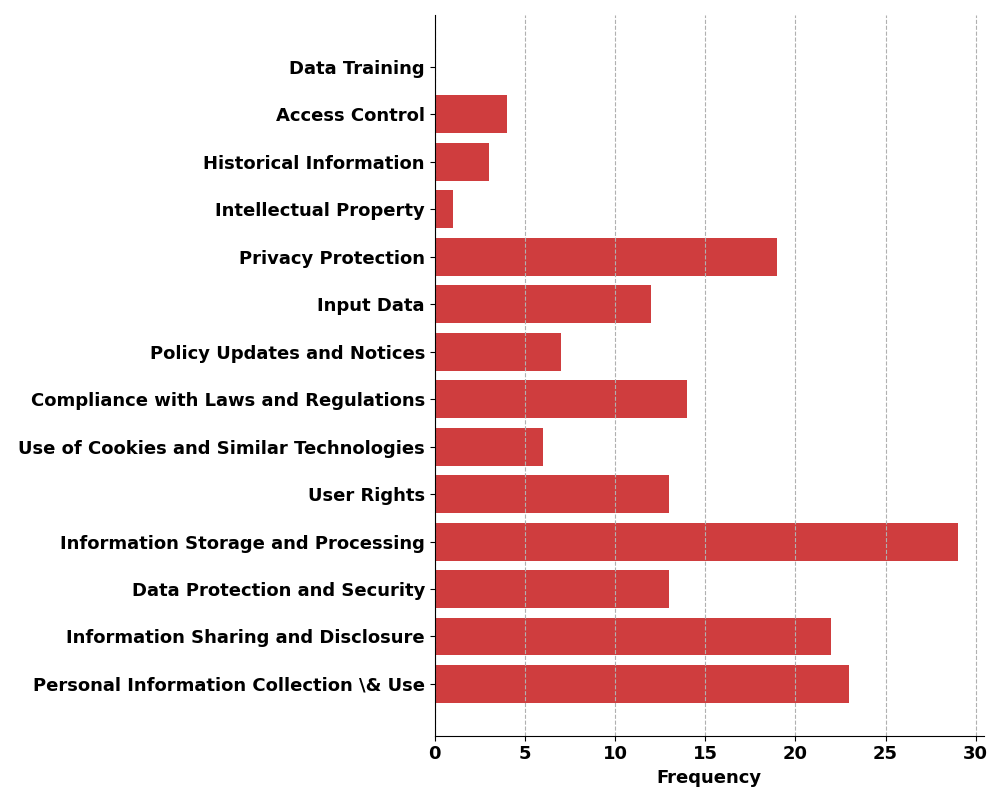}
    }

    \subfloat[]{
        \includegraphics[width=0.30\linewidth]{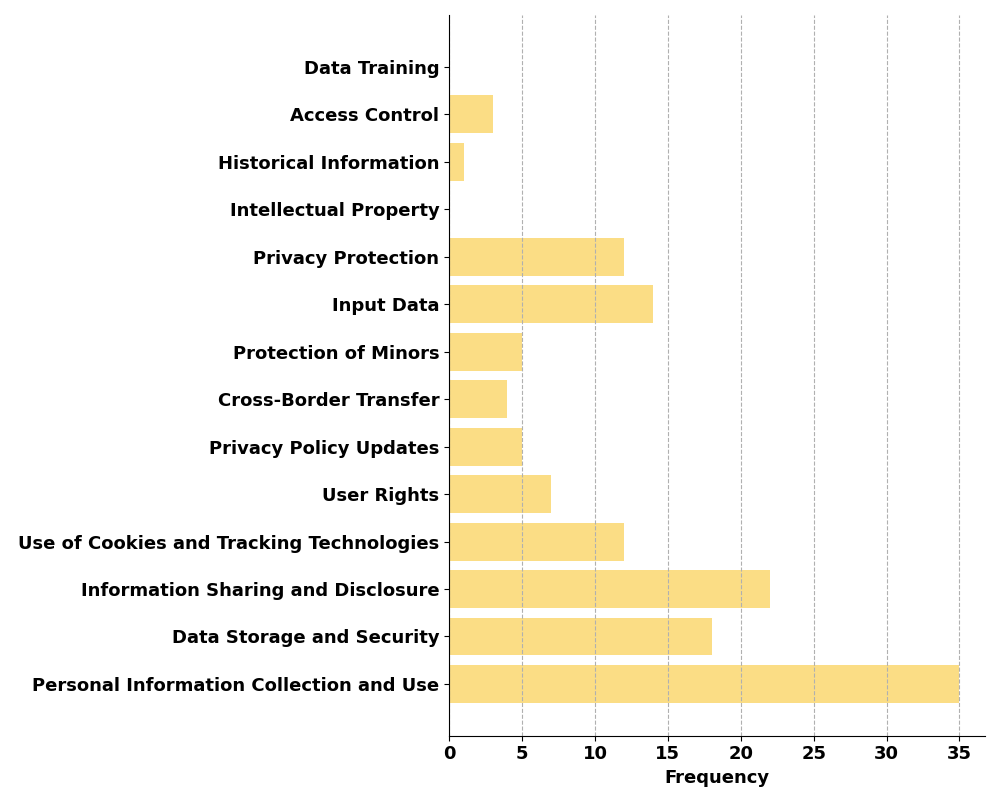}
    }
    \subfloat[]{
        \includegraphics[width=0.30\linewidth]{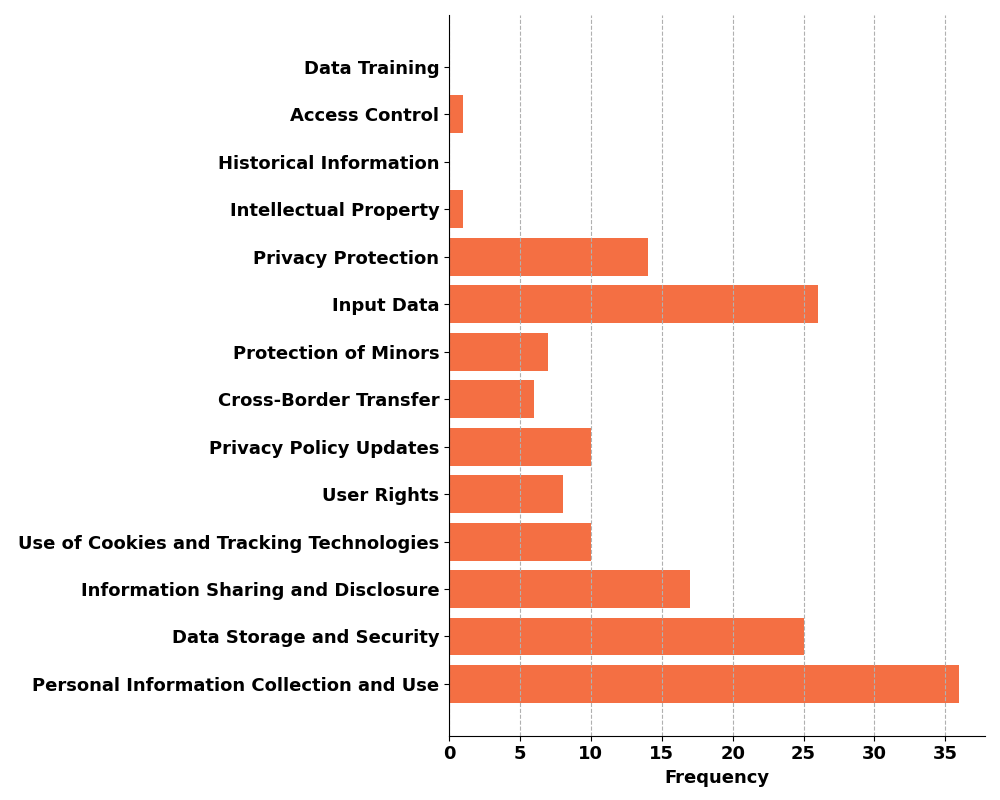}
    }
    \subfloat[]{
        \includegraphics[width=0.30\linewidth]{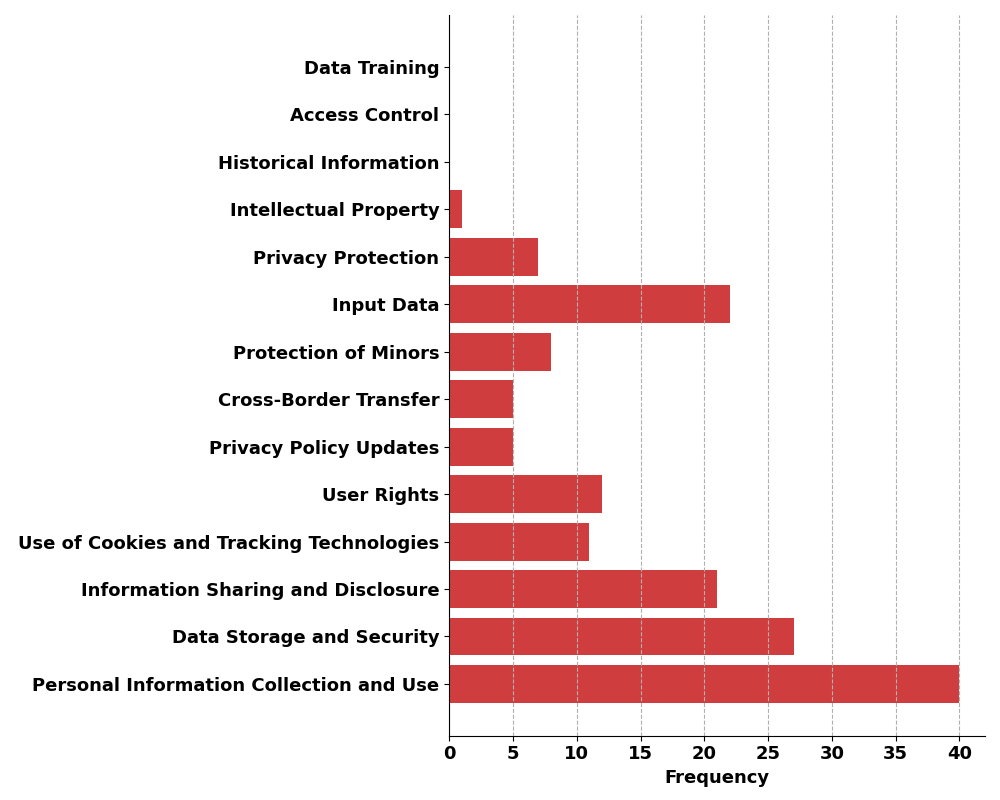}
    }
    \caption{The keywords and utterance of users for cursory reading and detailed reading. The x axis denoted the frequency of participants. (a), (b), (c) were extracted from the important information, memorized information, important information with reference during cursory reading of user agreement. Similarly, (d), (e), (f) were extracted from the important information, memorized information, important information with reference during detailed reading of user agreement. (g), (h), (i) were extracted from the important information, memorized information, important information with reference during cursory reading of privacy policy. (j), (k), (l) were extracted from the important information, memorized information, important information with reference during detailed reading of privacy policy.}
    \label{fig:cursory-detail}
\end{figure*}

Figure~\ref{fig:cursory-detail} demonstrated the main keywords each reading pass, question and document user read (generated through thematic analysis). We analyzed the results in detail and summarized the findings into the following 4 progressive insights.

\subsection{LLMs' User Agreement and Privacy Policy Content Structure}
\subsubsection{Content Structure}
Table~\ref{tbl:user_consent_content} and Table~\ref{tbl:privacy_policy_content} provided the content structure of user agreement and privacy policy of LLMs. The user agreement mainly focused on 73 different aspects, covering Personal Information Collection, Watermark, Intellectual Property Rights, etc. which was highly correlated with the LLMs' capability and service. Similarly, the privacy policy focused on 75 different aspects, with Data Source, Data Usage for Training, Data Protection, Disclosure as its main content. This was because for interaction with LLMs, participants needed to input personal information.  

\begin{table*}[!htbp]
\centering
    \begin{tabular}{c|c|c|c} 
    \hline
    \textbf{Term} & \textbf{Frequency} & \textbf{Term} & \textbf{Frequency} \\
    \hline
    Privacy Policy Changes & 12 & Appeals & 2\\
    Personal Information Collection & 12 & Legal Contract Basis and Principles & 2 \\
    Overview & 11 & Accuracy & 2\\
    Online Analysis and Cookies & 11 & Employment & 2\\
    Personal Information Retention & 11 & Sensitive Data Collection & 2 \\
    Contact Us & 11 & International Users & 2 \\
    Feedback and Correction & 10 & Services & 2\\
    Children & 10 & Authorized Agents & 2\\
    Disclosure of Personal Information & 9 & Account Creation & 2 \\
    Unsubscribe & 9 & Service Access Requests & 1\\
    Account Deletion & 9 & Intellectual Property & 1\\
    Collection Scenarios Not Requiring Consent & 9 & Illegal Use & 1\\
    User Rights & 9 & Risk Assumption & 1\\
    Security Measures & 9 & Severability & 1\\
    Deletion of Personal Information & 9 & User Responsibilities & 1 \\
    Personal Information Requests and Identity Verification & 9 & Source Demographics and Biases & 1\\
    Third-party Links & 9 & Partial Region Privacy Rights & 1\\
    Transfer & 8 & Data Declarations & 1\\
    Purposes of Personal Information Use & 7 & Data Protection Officer & 1\\
    Information Sharing & 7 & Effectiveness & 1 \\
    Exercising Rights & 6 & Responsible Document Collection & 1 \\
    Compliance and Protection & 6 & Multilingualism & 1 \\
    Security Measures and Disclaimers & 5 & Collective Litigation Waiver & 1 \\
    Information Storage Period & 5 & Legal Information Processing Documents & 1\\
    Aggregation or Anonymization of Information & 5 & Proactive Measures & 1\\
    Data Transfer & 4 & Do Not Track & 1 \\
    Rights and Obligations of Information Processing & 4 & Cookie Disabling & 1 \\
    Notification & 4 & Links to Other Websites & 1 \\
    Applicability & 4 & Advertising and Consumer Characteristics & 1 \\
    Definitions and Interpretations & 3 & Partial Region Disclosure of Information & 1 \\
    Training and Service Improvement & 3 & Illegal Content & 1 \\
    Marketing Communications & 3 & Telephone Communications & 1 \\
    Training Information and Development & 3 & Post-Deletion & 1 \\
    Region Restrictions & 3 & Arbitration & 1\\
    Response to Requests & 3 & Beneficiaries & 1 \\
    Appeals & 2 & Compensation & 1 \\
    Authorized Agents & 2 & Training Datasets & 1\\
    \hline
    \hline
    \end{tabular}
\caption{The Content Structure of Privacy Policies. The number after each entry represents the frequency of the term in the 12 LLMs' privacy policy surveyed.}
\label{tbl:user_consent_content}
\end{table*}

\begin{table*}[!htbp]
\centering
    \begin{tabular}{c|c|c|c} 
    \hline
\textbf{Term} & \textbf{Frequency} & \textbf{Term} & \textbf{Frequency} \\
\hline
Scope of Application & 12 & Who We Are / Definitions & 6 \\
Paid Accounts & 12 & Terms for Youths & 6 \\
Similarity of Content and Copyright & 12 & Accuracy & 6 \\
Software, Enterprises and Third-Party Services & 12 & Termination of Access & 6 \\
Limitation of Liability & 12 & Termination of Agreement & 6 \\
Various Countries & 12 & Personal Information Collection & 6 \\
Ownership of Content & 11 & Updates & 6 \\
Registration and Access: Registration & 11 & Feedback & 5 \\
Beneficiaries & 11 & User of User Account Information & 5 \\
Opting Out & 11 & Payments & 5 \\
Reservation of Rights & 11 & Export Controls & 4 \\
Service Termination & 11 & Complaints & 4 \\
Dos and Donts & 10 & User Behaviour Detection & 4 \\
Relationship between User Agreement and Privacy Policy & 10 & Suspension of Services & 4 \\
Disclaimer & 10 & User Rights & 4 \\
Registration and Access: Age Restrictions & 10 & Content Usage & 4 \\
Notifications & 10 & Open Source & 4 \\
Contact Information & 9 & User Obligations Under Agreement & 4 \\
User Rights under Agreement & 9 & Account Retrieval & 3 \\
Governing Law & 9 & Term Rights and Obligations Assignment & 3 \\
Content Provision and Liability & 9 & ICP Filling & 3 \\
Jury & 9 & Breach Handling & 3 \\
General Terms & 9 & User Information Provision Liability & 3 \\
User Account Obligations & 8 & Advertisements & 3 \\
Intellectual Property Rights & 8 & Commercial Confidentiality & 2 \\
Additional Terms for Google App Store & 8 & Additional Terms for Apple App Store & 2 \\
Communication Not Confidential & 8 & Severability & 2 \\
Third-party & 8 & Malicious Code and Security & 2 \\
Dispute Resolution & 8 & Watermarks & 2 \\
Changes to Terms & 8 & Management of Linked Accounts & 2 \\
User Responsibilities and Requirements & 7 & Data Deletion & 2 \\
Account Access & 7 & Bulk Arbitration & 2 \\
Subscriptions and Un-subscriptions & 7 & Data Protection & 2 \\
Account Security & 7 & Indemnification & 2 \\
Copyrights and Rights Trademark & 6 & Force Majeure and Unforeseen Events & 2 \\
    \hline
    \end{tabular}
\caption{The Content Structure of User Agreement.The number after each entry represents the frequency of the term in the 12 LLMs' privacy policy we surveyed}
\label{tbl:privacy_policy_content}
\end{table*}

\subsubsection{Difference Between LLMs and AIs' Privacy Policy and User Agreement}
We found through interview the most significant difference could be summarized as 6 categories: 1) Input Data (9/15, 60.0\%), 2) Privacy Protection (4/15, 26.7\%), 3) Intellectual Property (2/15, 13.3\%), 4) Historical Information (3/15, 20.0\%), 5) Access Control (3/15, 20.0\%), 6) Data Training (5/15, 33.3\%). These were highly correlated with LLMs' processing ability. Participants thought LLM-based products would ``reserve historical dialogue information'', had ``different regulations about intellectual property'' and needed to ``highlight privacy information protection''. In the following analysis, we would also focus on participants' concentration towards these difference.

\subsection{Ineffectiveness of Reading}
\subsubsection{Inattention of Users}

\begin{figure*}[htbp]
    \subfloat[Frequency to read LLM's privacy policy per year.]{
         \includegraphics[width=0.32\textwidth]{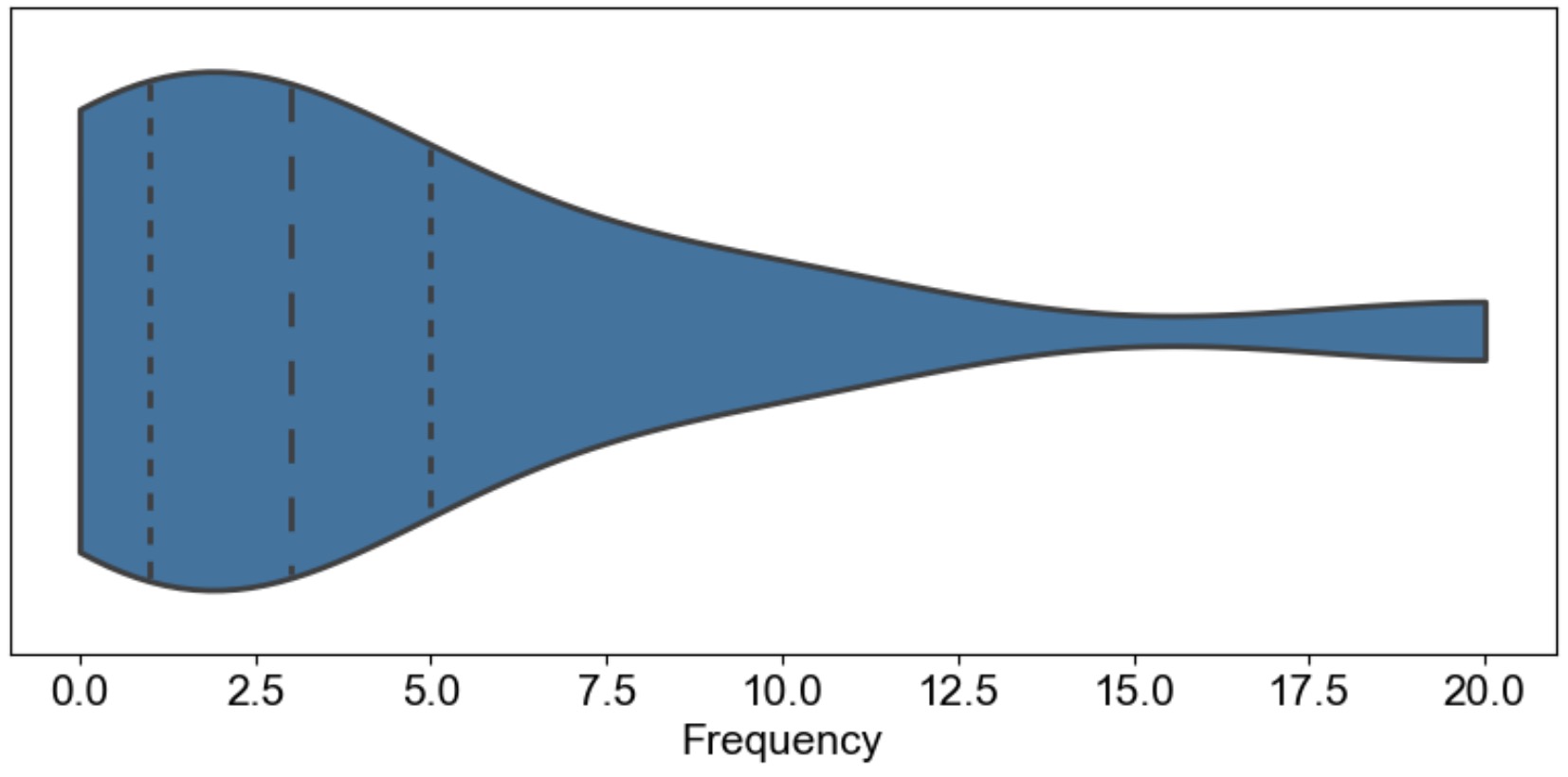}
    }
    \subfloat[Frequency to read AI's privacy policy per year.]{
        \includegraphics[width=0.32\textwidth]{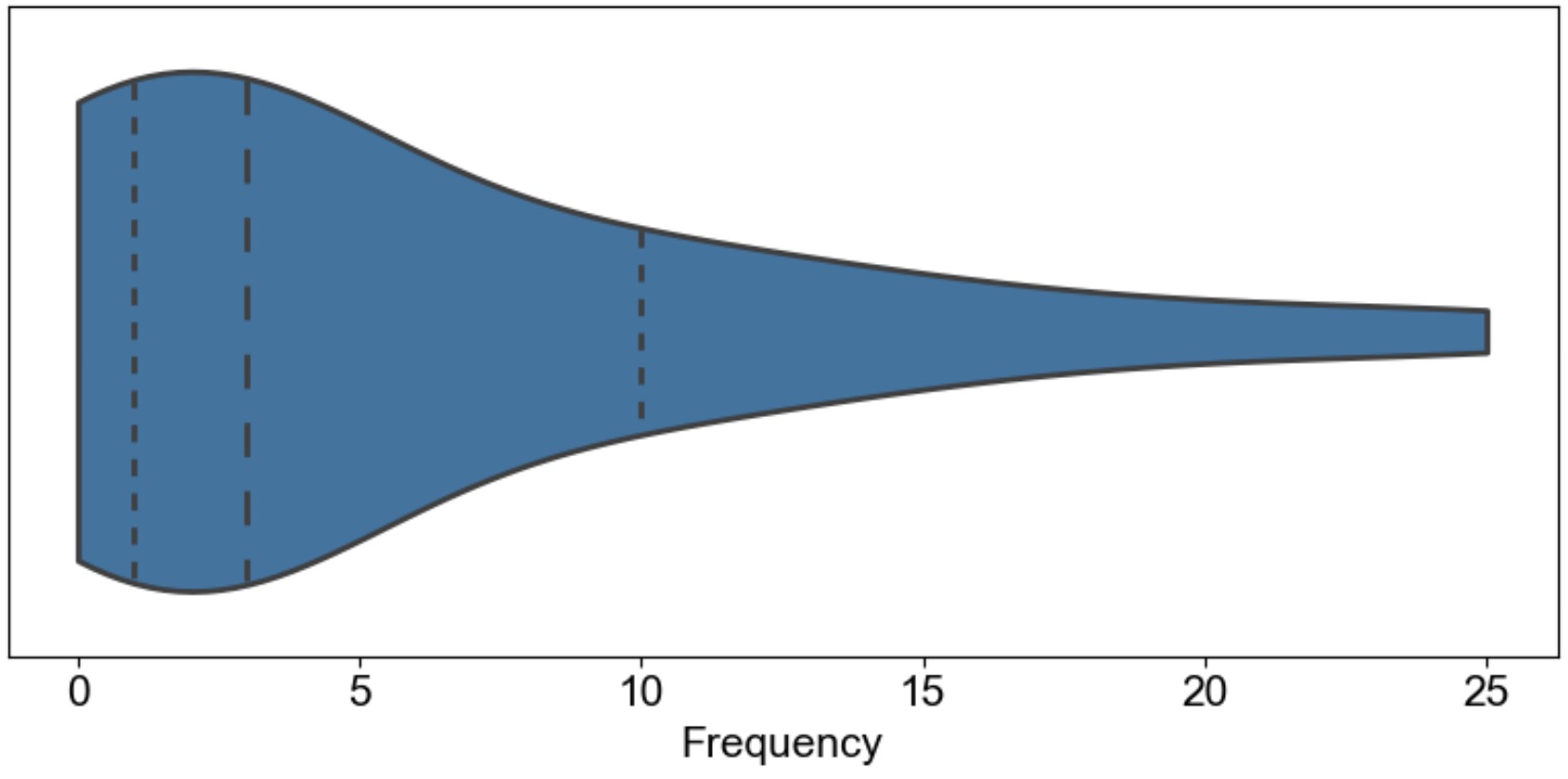}
    }
    \subfloat[Time to read AI's privacy policy each time.]{
        \includegraphics[width=0.32\textwidth]{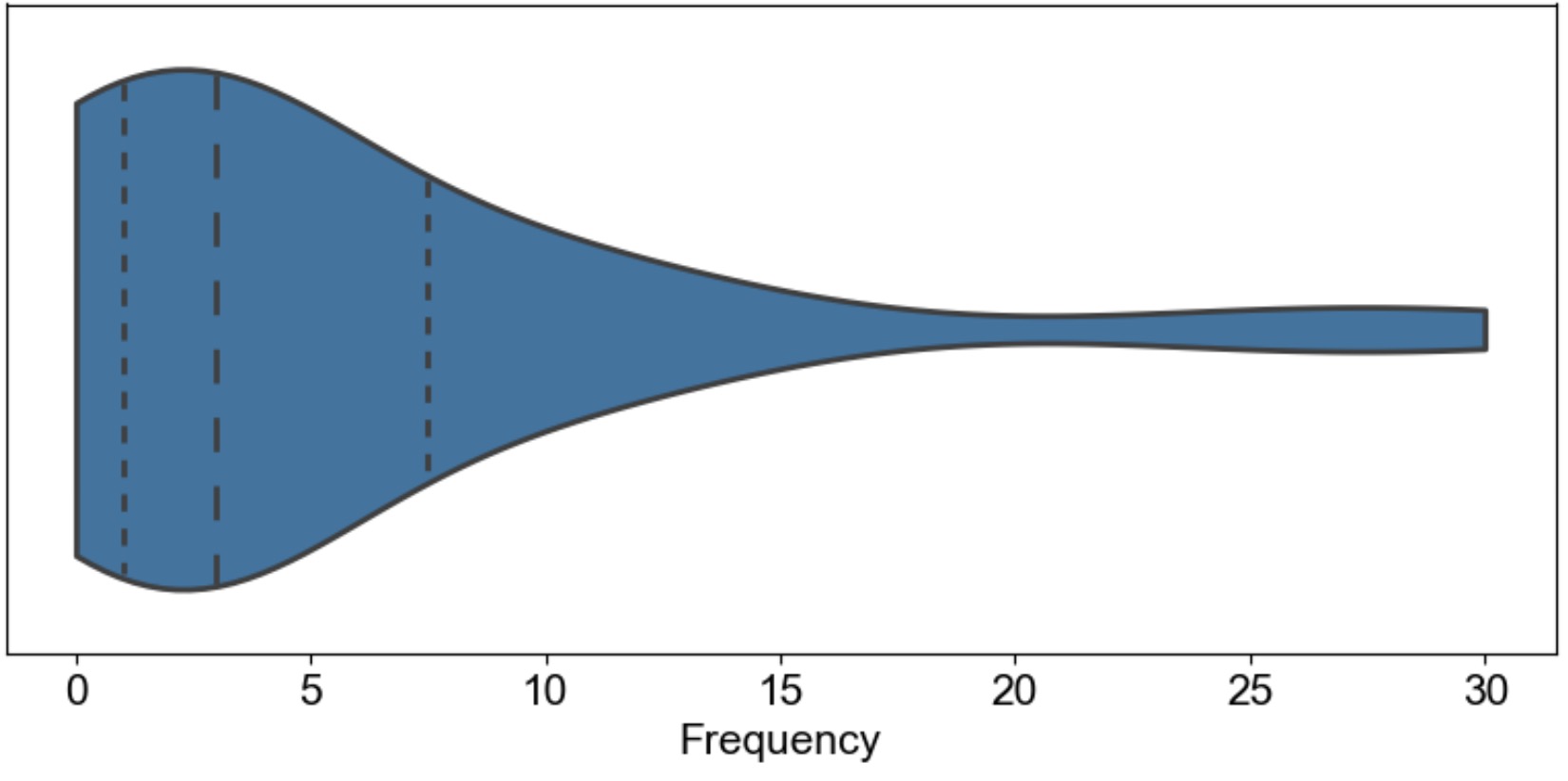}
    }
    \caption{Users' frequency to read the usage term and privacy policy.}
    \label{fig:inattention}
\end{figure*}

From Figure~\ref{fig:inattention}, we found users mostly would read the privacy policy only a few times a year and they would read the usage term and privacy policy only a few minutes each time (even 3/27 users said they would not even read usage term and privacy policy).

Analyzing deeper into the concentration points, from Figure~\ref{fig:cursory-detail}, we found participants would focus on legal compliance for both cursory reading and detailed reading. However, upon cursory reading, seldom would participants focus on privacy protection, which were the core of LLM's privacy policys. Besides, participants focused less on input data when cursory reading, which was the core difference between GPT and AI products' privacy policy. This implied the ineffectiveness upon a cursory reading and underscored the importance of a detailed reading. 

\begin{figure}
    \includegraphics[width=0.5\textwidth]{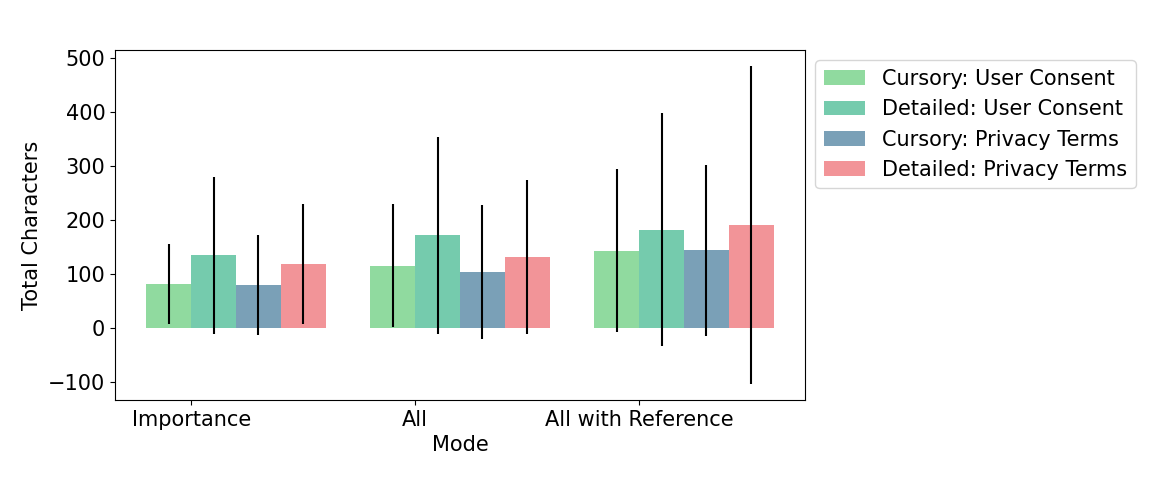}
    \caption{The characters of participants' utterance in each condition and each question. Errorbar showed one standard deviation.}
    \label{fig:plot_char}
\end{figure}

After detailed reading, the participants also remembered more content in total. Figure~\ref{fig:plot_char} showed the number of effective characters spoken by users after a cursory reading and a detailed reading, excluding duplicate words (prior to coding). Before reading the content, the users could only remember less than 100 characters (usually 50+ characters), however, given enough time to read, they would double the memory, which proved only after detailed reading could they remember some important information. In fact, significant difference was found between cursory reading and detailed reading (all $p < .05$).

This indicates that there is a lot of overlooked information by users during the cursory reading process, but they pay closer attention to it during the detailed reading process. This portion of information also deserves to be presented more effectively to attract users' attention.

\subsubsection{Core Attention of Users}

Although we found users lacked enough attention, we still wanted to analyze the core attention of users in order to provide better design guidance.

For the cursory reading, regarding user agreement, the most remembered information was ``service legality and reliability'', ``content and behaviour standards'' as well as ``commercial user and legal compliance''. These information was correlated with the normal usage of participants and may be correlated with fining. As LLMs' input and ouptut could potentially be applied to more usage, participants have more risk of misuse, resulting in more regulations to pay attention on. Through the interview, participants also agreed on the points were more correlated with products' normal functioning, thus they cared more. 

Regarding the privacy policy, the most cared points were ``information storage and processing'', ``input data'' and ``personal information collection \& use''. Participants also thought ``privacy protection'' and ``input data'' important although they need the reference of the original document.


After a detailed reading, users gained deeper understanding. During the cursory reading, participants had an average understanding of the user agreement and privacy policy at 3.22 and 3.48 respectively. Participants considered the content's effectiveness to be 4.07 and 3.93 respectively. However, after careful reading, users' understanding improved to 4.85 and 4.81 (user agreement: $Z = 0.0$, $p < .001$, privacy policy: $Z = 36.5$, $p < .05$). Besides, users' perception of the effectiveness of the user agreement and privacy policy increased to 4.69 and 4.76 respectively (user agreement: $Z = 8.5$, $p < .001$; privacy policy: $Z = 15.0$, $p < .001$).

\begin{figure}
    \centering
    \includegraphics[width=\linewidth]{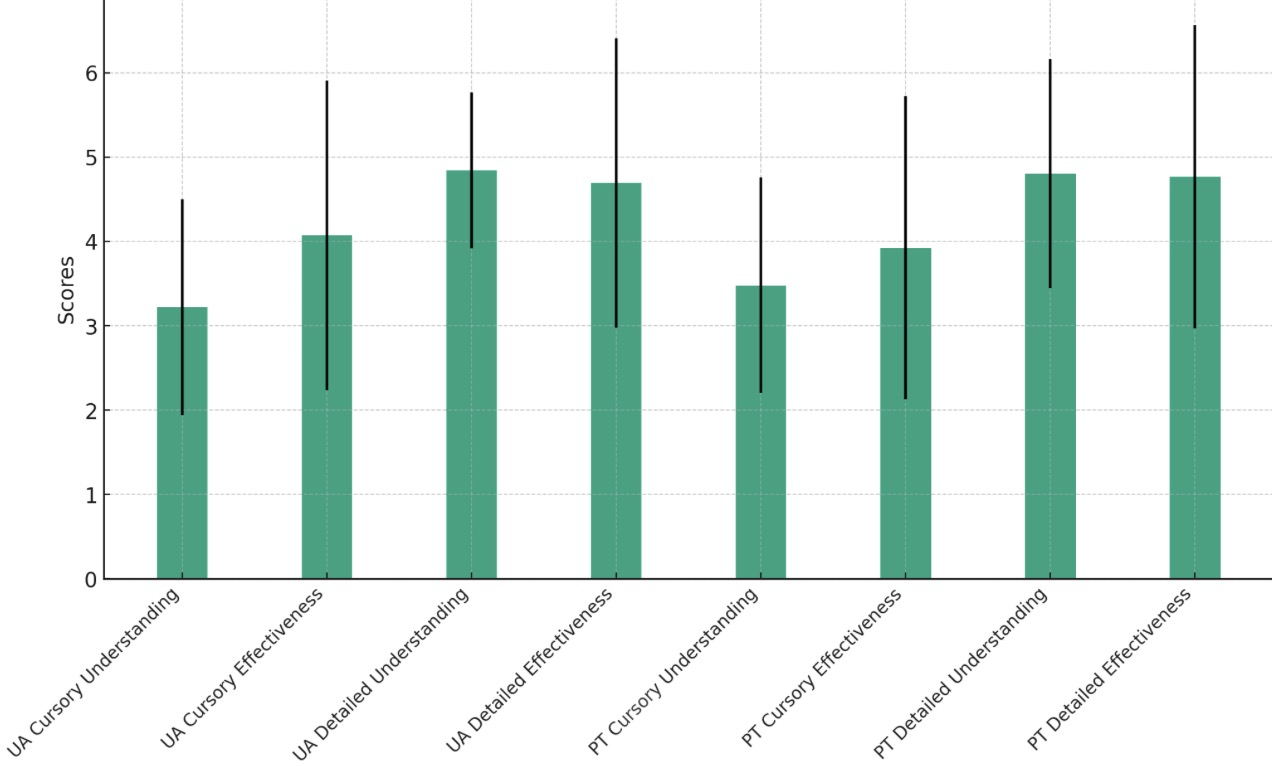}
    \caption{Participants' subjective ratings (7: most positive, 1: most negative). Errorbar indicated one standard deviation.}
    \label{fig:enter-label}
\end{figure}

Participants cared more about ``illegal activities and liability'', ``content creation and intellectual property'' as well as ``input data'' upon detailed reading for user agreement. Compared to cursory reading, a detailed reading process could further enhance users' attention on their own personal information. Besides, the core points upon detailed reading process focused more on detailed information (e.g., intellectual property, service changes), which reflected participants needed the detailed reading process to acquire more important information.

While for privacy policy, users cared more about ``usage of cookies``, ``privacy policy updates'' and ``cross-border transfer'' compared to cursory reading. This was similar to the results derived for user agreement, which proved the necessity for a detailed reading to acquire more information. During interview, participants also revealed that they would focus on more personal rights and obligation (e.g., user rights, use of cookies) rather than the information affecting normal usage (e.g., personal information collection). 

\subsection{Privacy Concerns Persisted with Full Understanding}

\subsubsection{Verification on the Integrity of the Content Read by the Users}



While we require users to thoroughly read the user agreement and privacy policy of selected LLM products, it's challenging to ascertain if users have meticulously read the entirety of the content. To address this, we asked a selected group of users what they thought they have read. During this question's answering, they could refer to the original content. This approach enabled us to determine whether the detailed reading was effective and thether participants read all the content. 

Our findings indicate that 75\% (6/8) of the users read all the content in full. For those who reported did not read everything, the unread proportion ranged no more than 30\%. The primary reasons for not reading certain sections in detail included: similarity in paragraph content, lack of user interest, and the perception of information as being common knowledge. From this, we concluded users who engage in detailed reading indeed pay close attention to the vast majority of the content.

\subsubsection{Gap Persisted upon Detailed Reading}\label{sec:gap_persisted}

Although in the previous section we found participants would focus on more previously omitted information, we found the gap persisted upon the detailed reading. 

First, even after a detailed reading process, seldom do users concern about the data training and access control of LLM-related products, which were the core difference between LLM-related products and AI-related products. Besides, even if most participants cared about the information sharing and disclosure, seldom do participants care about personal data deletion in the user agreement and privacy policy. This reflected the inattention of users even upon detailed reading. The phenomenon was caused by the fact that most participants ``read the content but could not remember them''. ``The content was too boring and tedious for me. ''

Second, we interviewed participants about whether after a detailed reading their privacy concerns were resolved. From the answers, 10/15 (66.7\%) had unresolved privacy concerns which concentrated at: 

\noindent \textbf{Users lack the right to choose}: 6/15 (40.0\%) of the participants lacked the right to choose whether ``they could control the data'' or whether ``they could choose whether to agree with the user agreement''. If they wanted to receive the service, ``sacrificing the personal information seems mandatory.'' For LLM-based products, this problem was stratified because all the conversations were in natural language, which meant the chance leaking personal information surged.

\noindent \textbf{Third-party use of information causes unease among users}: 2/15 (13.3\%) of the participants feared that third-party collection and usage of information would omit the privacy protection of the LLM-based product. In this setting, their personal information would also face potential disclosure. Because LLM-based products now usually supported searching \footnote{https://openai.com/blog/chatgpt-plugins/, accessed by 16th Jun, 2024}, plugins\footnote{https://platform.openai.com/docs/actions/introduction, accessed by 16th Jun, 2024}, retrieval augmentation \cite{lewis2020retrieval} and companies even would trade data, users' privacy data faced severe challenge.

\noindent \textbf{Distrust in the platform}: 4/15 (26.7\%) of the participants thought they could not trust in the privacy policy due to a lack of supervising. Besides, there was seldom the disclosure of the supervisors. Given that sometimes they have heard of privacy leakage matters, they could not trust in the platform. Further because the LLMs products contained novel technologies which participants did not know and understand, they could not easily trust these products without transparency.

\noindent \textbf{Explanations of information processing are insufficiently detailed}: 4/15 (26.7\%) of the participants thought the information processing was not explained clearly. ``How to distinguish personal sensitive information from normal information'' and ``What detailed ways would the company protect the users' information'' was not disclosed. This reduced participants' trust in the privacy protection of the products.
\subsection{How to Improve?}
In the previous sections, we first classified the difference between LLM-based products and AI-products' user agreement and privacy policy. Then we separately examined the inattention of cursory reading, the gap upon detailed reading and corresponding concentration to unveil the necessity of improving these policies. Thus, in this section we provided improvement goals and implementation sequentially. 

\begin{figure}
    \includegraphics[width=\linewidth]{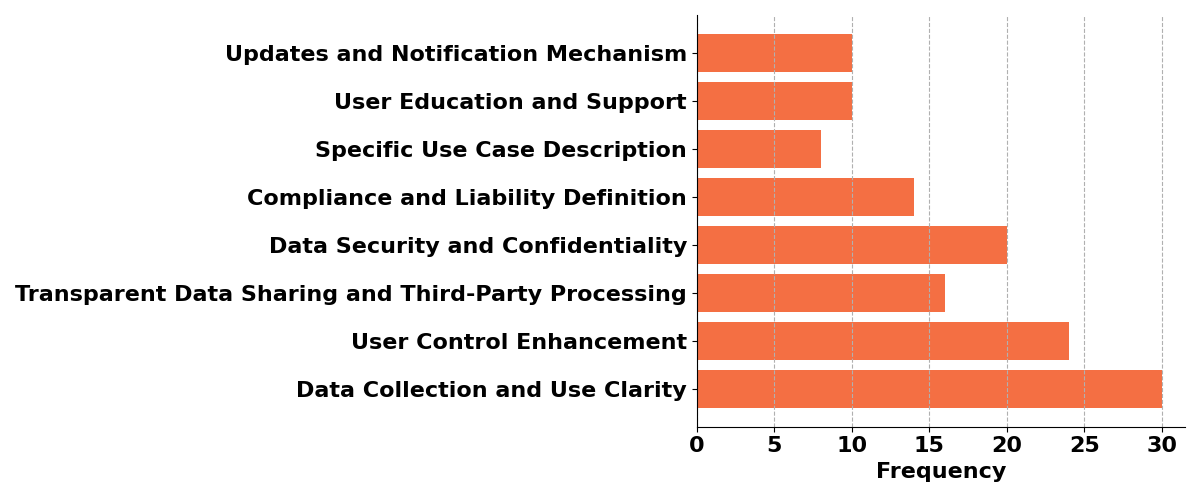}
    \caption{The frequency of participants' agreement with different design considerations.}
    \label{fig:improve_freq}
\end{figure}

\subsubsection{Implication 1: Clarity on Data Collection and Use}
As Figure~\ref{fig:improve_freq} showed, 30/42 (71.4\%) participants agreed with the need to clarify on data collection. In fact, also 20/42 (47.6\%) participants thought the manufacturers should use industry-standard encryption and anonymization technologies to product data during transmission and storage. ``I believe they should focus on privacy and inform users in detail where their information goes, which information will be read, and under what circumstances their rights to the information are protected.'' (P6) ``I would pay more attention to which information is collected, what it will be used for and who or which institutions it may be shared with.'' (P14)

\subsubsection{Implication 2: Enhanced User Control} 

24/42 (57.1\%) participants echoed the policies should enhance the users' control. Especially what actions could users do if they did not agree with the User Agreement should be designed and highlighted in the beginning. Otherwise, the problems in Section~\ref{sec:gap_persisted} persisted because users have no control of the private data. ``It's important to clarify how to opt-put if one decides not to use the product. This opt-out process must be clearly written and should be positioned relatively at the beginning.'' (P18)

\subsubsection{Implication 3: Transparent Data Sharing and Third-Party Processing}

16/42 (38.1\%) participants hoped the policies could disclose data sharing and third-party processing details and made them transparent. This could also respond to the concerns in Section~\ref{sec:gap_persisted} that participants did not know the data processing and sharing process to third-party institutions. Nor could they guarantee the information privacy.
``For technical aspects, it is essential to clearly explain what technologies are being used to avoid any disputes in this regard.'' (P18) ``Clearly list what users cannot do and the consequences of those actions, and try to describe transparently under what circumstances privacy might not be `fully protected.' '' (P34) 

\subsubsection{Implication 4: User Education and Support}

10/42 (23.8\%) participants appealed for better user education and support towards the product as well as the privacy protection. They hoped interactive visualization (e.g., highlight) and education technologies could be used to help them better use the product.
``I can't understand some parts like international information transmission even reading carefully. I hope they are not presented in text only, or could be explained with animations to highlight key points.'' (P9) ``I hoped they could adopt interactive explanation interface to indicate where information might be collected and how information will be used.''

\section{General Discussions}
\subsection{LLM-related Products versus AI Products}
LLM-related products are distinct from AI products in several ways: firstly, LLMs operate on a large scale, which necessitates the frequent utilization of user-generated data. This added new potential harm to users when assessing privacy risk and viewing privacy policy \cite{carlini2022quantifying,carlini2021extracting,zhang2023counterfactual}. According to privacy risk's definition \cite{kaplan1981quantitative}, privacy calculus theory \cite{laufer1977privacy} and technology acceptance model (TAM) \cite{davis1985technology}, users' acceptance of LLM-related products would be largely affected by these potential misuse. Thus, understanding the reading behavior or privacy policy, which clearly stated the potential use of users' data, is essential.

Besides, the mental model of users interacting with LLMs is different from traditional AIs \cite{zhang2024s}. Users hold a in-comprehensive mental model and may be more willing to disclose their private information due to a higher anthropomorphism \cite{ischen2020privacy,kim2012anthropomorphism} and the benefit of functionalities \cite{zhang2024s}. This poses new privacy risk which urges participants to read and understand privacy policy. 

Lastly, most text-based and image-based LLMs collected different information from traditional AIs \cite{zhang2024s}, which also extended on the information during human-human interaction on online social platform \cite{richthammer2014taxonomy,richthammer2013taxonomy}. These changes, officially, were only reflected in the privacy policy \cite{}. Thus, addressing the reading of privacy policy is the first step to ensure users' control over their data and mitigate the privacy risk associated with LLM-based products.

\subsection{Users' Inattention of Reading}

Various past work revealed the fact that users did not attentively read privacy policy. However, the LLMs lacked law's regulation now, whose privacy policy were less structured and may comprise more content (e.g., information needed to collect, more usage). This worsened the comprehensiveness of participants' reading. Different from the former work which put the focus on the length of privacy policy \cite{meier2020shorter,mcdonald2008cost}, the expectation on the form of the privacy policy \cite{youn2009determinants,xu2011information}, we focused on the factor: the time spent on reading the privacy policy. We revealed that even given enough time to read the privacy policy, users were still lacked behind in understanding LLMs' privacy policies. We advocate for better illustration and designs for facilitate reading (e.g., novel forms of privacy labels \cite{johansen2022multidisciplinary,zhang2022usable}, plugins \cite{cui2019ppsb}).

\section{Ethical Considerations}
We acknowledged the potential ethical problems and meticulously designed the experiment process. Besides, we got the approval of our Institutional Review Board (IRB). We introduced our effort in reducing potential privacy concerns following the guidelines of Menlo Report \cite{bailey2012menlo} and Belmont Report \cite{beauchamp2008belmont}. In user study, we followed the standard ethical research procedures~\cite{zhang2022ethics} to ensure the confidentiality of information provided by participants. We did not open-source any of the data, nor shared that for any other research or commercial use. Participants in the user study have the right to stop the experiment and erase their data at any time for any reasons. Before the experiment, we would also show the participants the User Agreement before progressing. For analysis, all the personal information of participants was properly anonymous and also encrypted. We stored it on a personal computer without uploading any of the data on the cloud or copying the data out of the computer. Our study aimed to benefit the future privacy policies' design and in fact would also increase participants' privacy protection awareness.

\section{Limitation and Future Work}
We pointed out the potential limitations in our research, which we also saw as future improvement directions.
First, our selection of the products as well as the participants had potential limitation. As a first attempt, we chose the text-based LLMs products to evaluate, which was the most famous and widely-adopted type. The future research direction could be extended to voice or image-based products which had other problems. Besides, our participants were limited to Chinese youths, this was partially because we needed the participants to have used LLM-based products. The future research direction could be extended to experiments grounded in other countries or regions.
Second, we mainly investigated users' opinions towards LLM-based products without grounding the privacy policy knowledge. In fact, participants gave the most direct feedback and evaluation of the products and determined whether to use LLM-based products, thus their opinion was the most important. However, if the knowledge from legal and law perspective could be integrated, the analysis of privacy policies could be more well-rounded.

\section{Conclusions}
LLM-based products was surging in the current society, with nearly billions of users adopting LLM-based products for content generation, text editing, searching, etc. However, the effectiveness of their privacy policy was unexplored till now. Because users adopt natural language to prompt LLM-based products, which potentially would leak sensitive user data without awareness, designing effective privacy policy was important. Thus, we conducted the first study (N=42) to investigate the difference, the ineffectiveness and the potential improvement direction of LLM-based products. We found 1) the core difference of GPT-based products and AI-based products lied in the sensitive information processing and problems (e.g., intellectual property) of users' input. 2) users did not pay enough attention to the privacy policy and omit important detailed points. 3) users had additional concerns even if they carefully read policies. Based on the former conclusions, we proposed 4 design implications regarding the transparency, the details of the information processing towards manufacturers and content providers. 

{\footnotesize \bibliographystyle{acm}
\bibliography{sample}}

\appendix

\section{Selection of LLMs}
We provided our selection of LLMs according to the users and their targeted area. Because we wanted to know whether the privacy perception of the end users, we only consider general usage LLMs.

\begin{table*}[!htbp]
    \centering
    \caption{Usage of SPA, the collected information and their frequency.}
    \begin{tabularx}{\textwidth}{|p{75px}|p{70px}|p{33px}|p{35px}|p{25px}|p{18px}|p{30px}|p{30px}|p{30px}|p{35px}|}
    \hline
        Name            & Company                       & Privacy Policy  & Terms of Use  & Login & Use & Manual Search & Tokens Terms & Tokens Policy & Country / Region \\ \hline
        Wen Xin Yi Yan  & Baidu                         & $\checkmark$  & $\checkmark$  & $\checkmark$  & $\checkmark$  & $\checkmark$ & 7,030  & 7,873     & Beijing \\ 
        Qing Yan        & Zhipu                         & $\checkmark$  & $\checkmark$  &  $\checkmark$ & $\checkmark$  & $\checkmark$ & 8,346  & 8,976     & Beijing \\
        Bai Chuan       & Baichuan                      & $\checkmark$  & $\checkmark$  &  $\checkmark$ & $\checkmark$  & $\checkmark$ & 8,071  & 7,490     & Beijing \\
        Shu Sheng       & Shanghai Ai Lab               & $\checkmark$  & $\checkmark$  &  $\times$     & $\times$      & $\checkmark$ & 9,589  & 11,515    & Shanghai \\ 
        Xing Huo        & Keda Xunfei                   & $\checkmark$  & $\checkmark$  & $\checkmark$  & $\checkmark$  & $\checkmark$ & 6,191  & 13,719    & Hefei \\
        Ri Ri Xin       & Sensetime                     & $\checkmark$  & $\checkmark$  & $\checkmark$  & $\checkmark$  & $\checkmark$ & 22,266 & 5,933     & Shanghai \\
        Dou Bao         & Chun tian Zhi yun (tiktok)    & $\checkmark$  & $\checkmark$  & $\checkmark$  & $\times$      & $\checkmark$ & 11,192 & 8,061     & Beijing \\
        Zi Dong Tai Chu & Chinese Academy               & $\checkmark$  & $\checkmark$  & $\checkmark$  & $\checkmark$  & $\checkmark$ & 11,265 & 2,085     & Hefei \\ 
        Tong Yi Qian Wen & Ali                          & $\checkmark$  & $\checkmark$  & $\checkmark$  & $\checkmark$  & $\checkmark$ & 12,892* & 12.892* & Beijing \\ 
        Wu Dao           & BAAI                         & $\checkmark$  & $\checkmark$  & $\times$      & $\times$      & $\times$     & --     & --        & Beijing \\ 
        Hun Yuan        & Tengxun                       & $\checkmark$  & $\checkmark$  & $\checkmark$  & $\checkmark$  & $\checkmark$ & 9,295  & 5,845     & Beijing \\ 
        ChatGPT         & OPENAI                        & $\checkmark$  & $\checkmark$  & $\checkmark$  & $\checkmark$  & $\checkmark$ & 3,314  & 3,146     & America \\ 
        Coral           & Cohere                        & $\checkmark$  & $\checkmark$  & $\checkmark$  & $\times$      & $\checkmark$ & 8,191  & 2,626     & America \\
        PI              & Inflection AI                 & $\checkmark$  & $\checkmark$  & $\checkmark$  & $\checkmark$  & $\checkmark$ & 3,339  & 7,657     & America \\ \hline
    \end{tabularx}
\label{llm_selection}
\end{table*}

\section{Participants' Demographics Background}
In this section, we provided the detailed demographics background as shown in Table~\ref{formal_background}.
\begin{table*}[!htbp]
    \centering
    \caption{The demographics of the participants}
    \begin{tabularx}{\textwidth}{|p{240px}|p{150px}|X|}
    \hline
        Class & Level & Number \\ \hline
        \multirow{2}{80px}{Age} & Male &  \\
          & Female &  \\ 
          \hline
        \multirow{4}{*}{Number of times using LLM products annually} 
         & 0-50 &  15 \\ 
         & 50-100 &  4 \\
         & 100-1000 &  18 \\
         & 1000+ &  2  \\
          \hline         
          \multirow{4}{*}{Number of times reviewing LLM's privacy policy annually} 
         & Hardly ever read &  7 \\ 
         & 1-5 &   24\\
         & 5-20 &   8\\
         & 20+ &  0  \\
          \hline
        \multirow{3}{*}{Educational background} 
         & Bachelor & 20 \\
         & Master & 11 \\
         & Ph.D. & 8 \\ 
         \hline
    \end{tabularx}
    \label{formal_background}
\end{table*}

\section{Participants’Attention on Detailed Reading }
We provide the main content that users pay attention to when reading in detail, and summarize these content into keywords. The users' attention 
on privacy policies mainly focused on 22 different aspects, while those regarding user agreements are 30 aspects. The results are shown in Table ~\ref{frequency_of_privacy_policy} and Table ~\ref{frequency_of_user_agreement}. The total number of users surveyed is 15.

\begin{table*}[!htbp]
\centering
\caption{Keywords of Users' Attention on Privacy Policies and Frequencies}
    \begin{tabular}{|c|c|c|c|} 
    \hline
        Region Restrictions & 1 & Personal Information Collection & 6 \\
        Online Analysis and Cookies & 2 & Purposes of Personal Information Use & 3 \\
        Data Transfer & 1 & Personal Information Retention & 6 \\
        Security Measures & 3 & Children & 5 \\ 
        Governing Law & 3 & User Rights & 2 \\ 
        Security Measures & 2 & Disclosure of Personal Information & 3 \\
        Risk Assumption & 1 & Third-party Links & 4 \\ 
        Aggregation or Anonymization of Information & 3 & Deletion of Personal Information & 4 \\
        Account Deletion & 2 & Payments & 1 \\
        Data Protection Officer & 1 & Personal Information Requests and Identity Verification & 1 \\
        Information Sharing & 1 & Collection Scenarios Not Requiring Consent & 3 \\ \hline
        \end{tabular}
    \label{frequency_of_privacy_policy}
\end{table*}

\begin{table*}[!htbp]
\centering
\caption{Keywords of Users' Attention on User Agreement and Frequencies}
    \begin{tabular}{|c|c|c|c|} 
    \hline
        Registration and Access & 5 & Account Access & 1 \\
        Governing Law & 7 & Payments & 3 \\
        Disclaimer & 4 & Dos and Donts & 6 \\
        Content Usage & 2 & Subscriptions and Un-subscriptions & 1 \\
        Personal Information Collection & 1 & User Responsibilities and Requirements & 4 \\
        Registration and Access: Age Restrictions & 6 & Account Security & 2 \\
        Complaints & 1 & User Rights & 2 \\
        Content Provision and Liability & 1 & Changes to Terms & 2 \\ 
        Intellectual Property Rights & 3 & Reservation of Rights & 2 \\ 
        User Account Obligations & 2 & Updates & 1 \\
        Software, Enterprises and Third-Party Services & 1 & Ownership of Content & 2 \\
        Limitation of Liability & 1 & Data Deletion & 1 \\
        Force Majeure and Unforeseen Events & 1 & Malicious Code and Security & 1 \\
        Commercial Confidentiality & 1 & User of User Account Information & 3 \\
        User Information Provision Liability & 1 & Various Countries & 1 \\ \hline
        \end{tabular}
    \label{frequency_of_user_agreement}
\end{table*}

\end{document}